\documentclass[toc]{PoS}

\usepackage{amsmath,amssymb}
\usepackage{graphicx}
\usepackage{feynmp}
\usepackage{slashed}
\usepackage{url}

\def\eg{{\sl e.g. \,}}
\def\ie{{\sl i.e. \,}}
\newcommand{\Mpl}{M_\text{Planck}}
\newcommand{\Mstar}{M_\star}
\newcommand{\Mrs}{M_\text{RS}}
\newcommand{\mkk}{m_\text{KK}}
\newcommand{\p}{\partial}
\newcommand{\mat}{\mathcal{M}}
\newcommand{\lag}{\mathcal{L}}
\newcommand{\amp}{\mathcal{A}}
\newcommand{\s}{\mathcal{S}}
\newcommand{\ope}{\mathcal{O}}
\newcommand{\cutoff}{\Lambda_\text{cutoff}}
\newcommand{\mtrans}{\Lambda_T}
\newcommand{\tev}{\text{TeV}}
\newcommand{\gev}{\text{GeV}}

\def\openone{\leavevmode\hbox{\small1\kern-3.8pt\normalsize1}}%

\title{Extra Dimensions and their Ultraviolet Completion}

\ShortTitle{UV-Complete Extra Dimensions}

\author{Erik Gerwick \\
        SUPA, School of Physics and Astronomy, 
        University of Edinburgh, Scotland \\
        E-mail: \email{E.Gerwick@sms.ed.ac.uk}}

\author{\speaker{Tilman Plehn}\\
        Institut f\"ur Theoretische Physik, 
        Universit\"at Heidelberg, Germany \\
        E-mail: \email{plehn@uni-heidelberg.de}}

\abstract{
  Large extra dimensions are one of the constructions addressing the
  hierarchy problem of the Standard Model. Their main theoretical and
  phenomenological challenge is that already predicting LHC effects
  requires an ultraviolet completion of TeV-scale gravity. In these
  lecture notes we first give a basic introduction into TeV-scale
  gravity models and their collider effects and then discuss possible
  ultraviolet completions, like string theory and fixed-point gravity.}

\FullConference{Workshop on Continuum and Lattice Approaches to Quantum Gravity\\
                 September 17-19 2008\\
                 Brighton, University of Sussex, United Kingdom}

\begin{document}

\newpage

\begin{fmffile}{feyn}

Many years of experimental tests have lead us to accept the Standard
Model as the effective theory valid at and below the weak energy
scale. This includes its range in terms of direct particle searches as
well as high-precision tests of quantum effects, both of them related
by the renormalizability of gauge theories. However, the crucial
ingredient of electroweak symmetry breaking is not yet understood, \ie
we still have not seen a fundamental scalar Higgs boson or any kind of
indication of strong interactions breaking the electroweak $SU(2)_L
\times U(1)_Y$ symmetry to the observed electromagnetic
$U(1)_Q$. While the search for some variation of a simple fundamental
Higgs scalar has been the major motivation in experimental searches as
well as theory model building for at least the past 25 years there is
a little more to it.  To date we see four serious problems with our
Standard Model
\begin{enumerate}
\item experimentally, it does not include dark matter, even though
  generically dark matter could be explained by a stable weak-scale
  particle with typical weak-scale Standard Model couplings~\cite{dm}.
\item theoretically, a truly fundamental quantum theory including
  masses for the $W$ and $Z$ gauge bosons should either include a
  Higgs boson or an additional strong interaction with its appropriate
  resonances. All we know to date is that a light Higgs scalar is
  consistent with electroweak precision data~\cite{lepewwg}.
\item if the Higgs boson is a fundamental scalar, its mass has to be
  protected. Otherwise, quantum corrections would betray the
  underlying principle of fundamental gauge theories and force us to
  order by order fine tune a counter term to stabilize the fragile
  Higgs mass --- the hierarchy problem~\cite{hierarchy}. Decoupling a
  corresponding new-physics sector from the electroweak precision data
  mentioned in point~(2) can be achieved with a discrete symmetry
  which in passing introduces a stable dark matter particle as
  required by point~(1).
\item gravity is not included in this picture of particle physics,
  even though we know that it includes the remaining fourth
  fundamental force between particles.
\end{enumerate}
Note that this is of course not a complete list of problems in
fundamental physics, which would have to include the cosmological
constant, the baryon asymmetry of the Universe, or the absence of
gravitational waves. This list simply includes issues which might well
be solved by TeV-scale new physics.\bigskip

On the other hand, this list makes it obvious that Higgs searches, or
searches for the mechanism of electroweak symmetry breaking, cannot be
separated from searches for TeV-scale new physics. Both are different
sides of the same medal. Proof that all four problems can indeed be
linked together is given by supersymmetry: by roughly doubling the
Standard Model's particle spectrum above the TeV scale it provides a
dark matter candidate, radiatively breaks electroweak symmetry,
stabilizes the Higgs mass, allows for a perturbative extrapolation to
high energies (including a grand unified theory) and links the
Standard Model to a local theory of gravity. The problem is that even
the minimal supersymmetric Standard Model can be viewed as more on the
elaborate than on the minimal side. Instead, we can ask the question:
how far can we get in solving as many of the above issues with as
little extra input as possible?\bigskip

Tackling the last of the problems listed above we run into a
fundamental problem of field theory --- we know from the classical
theory that the gravitational coupling carries a mass dimension, which
means we cannot quantize it in a perturbatively renormalizable
manner. What we can do is explicitly exclude the possibly dangerous
high-energy regime and treat gravity as an effective field theory, \ie
a theory with a built-in cutoff scale which should nevertheless
describe low-energy observables well. In Sections~\ref{sec:add} and
\ref{sec:rs} we will construct two such effective theories of
extra-dimensional gravity valid up to LHC energies and show the
limitations of this approach. In Section~\ref{sec:uv} we will compute
the same observables based on two ultraviolet completions of gravity
which cure the poor ultraviolet behavior of the effective theory of
gravity.

%%%%%%%%%%%%%%%%%%%%%%%%%%%%%%%%%%%%%%%%%%%%%%%%%%%%%%%%%%%%%%%%%%%%%%%
\section{Flat extra dimensions}
\label{sec:add}

One answer to this question is given by large extra
dimensions~\cite{add,tasi_add}. This model has the most important
feature that it does not introduce any additional states, which in
turn means that we would have to invoke some other mechanism to
explain dark matter. But as we will see below, it does successfully
tackle the three other problems and might even offer an explanation
for the small cosmological constant.\bigskip

Initially, large extra dimensions were suggested as an explanation for
the observed hierarchy between the electroweak and Planck~\cite{add}
or GUT~\cite{unification} scales while allowing the Higgs mass to
remain comfortably at a mass around $100\;\gev$. Such models with
large (compared to the Planck length) and flat extra dimensions are
referred to as ADD models. Their basis is a low fundamental Planck
scale ($\Mstar\sim \tev$) which also locates the onset of quantum
gravitational effects. This new scale serves as the ultraviolet cutoff
in the loop contributions to the renormalized Higgs mass, which limits
the size of quadratic quantum corrections.  This construct appears to
be in clear contradiction to all 4-dimensional data which determines
the Planck mass from Newton's constant $G_N\sim 1/\Mpl^2$, describing
the force on an object in a gravitational field.  The ADD model solves
this apparent contradiction by deriving the observed value of $\Mpl$
from the fundamental Planck mass $\Mstar$ and a a particular geometry
of space-time.

At the classical level we can see how this occurs in a universe with
extra spatial dimensions.  The Einstein-Hilbert action in any number of
$(4+n)$ dimensions is given as
\begin{equation}
 S_\text{bulk} = - \frac{1}{2} \int d^{4+n} x \; 
 \sqrt{-g^{(4+n)}} \; \Mstar^{n+2} \; R^{(4+n)}.
 \label{eq:eh_action}	
\end{equation}
We denote 4-dimensional space-time coordinates with Greek indices
$\mu,\nu,\alpha=0,1,2,3$ and extra dimensional coordinates with lower
case Roman letters $a,b,c=5,6,7,\cdots n$.  These are unified to
capital Roman letters $M,N,L=0,1,2,3,5, \cdots n$.  The 4-dimensional
coordinates we write as $x_\mu$, while the $(4+n)$-dimensional 
coordinates are $y_a$, such that $z_M=x_\mu+y_a$.\bigskip

The Einstein Hilbert action has a number of interesting features:
first of all, the propagating degrees of freedom are carried
exclusively by the the metric $g_{MN}$.  To act as a metric in the
conventional sense (most notably connecting vectors to form inner
products) it is symmetric, and to produce the correct mass dimensions,
$g_{MN}$ must be dimensionless. The mass dimension of the Ricci scalar
is independent of the underlying space-time dimension. This is merely
the statement that the dimensionality of $R$ is completely determined
by derivatives and not fields.  The action eq.(\ref{eq:eh_action})
enjoys full $(4+n)$-dimensional invariance under general coordinate
transformation with an arbitrary parameter $\xi_M(z)$
\begin{equation}
z_M\rightarrow z_M+\xi_M(z),
\end{equation}
where the induced variation in the metric is
\begin{equation}
\delta g_{AB}=\p_A \xi^M g_{MB}+\p_B \xi^M g_{MA}+\xi^M \p_M g_{AB}.
\label{eq:metvariation}
\end{equation}
The ADD model breaks this symmetry explicitly by treating the
4-dimensional space $x_\mu$ and the $n$-dimensional space $y_i$
differently. Coordinate transformation can no longer mix these
components.  The requirements on the space described by metric
$g_{ij}$ are
\begin{itemize}	
\item[--] \emph{spatial}: the signature for the $n$ extra dimensions
  is $(-1,-1, \cdots) $.
\item[--] \emph{separable}: the extra dimensions must be orthogonal to
  the brane so that the measure $d^{4+n}z$ is well defined.  In other
  words, the metric decomposes as a product space $g^{(4+n)} = g^{(4)}
  \otimes g^{(n)}$.
\item[--] \emph{flat}: the dimensions must be flat so that they can be
  integrated out explicitly in the action. In standard gravity the
  same is true unless sources induce $T_{ij}\ne 0$. We therefore
  restrict matter to the $y_i=0$ brane:
\begin{equation}
T_{AB}(x; y) = \eta^\mu_A \, \eta^\nu_B \, T_{\mu\nu}(x) \, \delta^{(n)}(y)
             = \left( \begin{array}{cc}
                      T_{\mu\nu}(x) \, \delta^{(n)}(y) & 0 \\ 0 & 0 
                      \end{array} \right).
\label{eq:conEMT}  			
\end{equation}
The assumption of an infinitely thin brane for our 4-dimensional world
might have to be weakened to generate realistic higher-dimensional
operators for flavor physics or proton
decay~\cite{fat_brane,discrete_gauge}. Einstein's equation purely in
the extra dimensions
\begin{equation}
R_{jk} - \frac{1}{n+2} g_{jk} R = 0. 
\end{equation}
contracted with $g^{jk}$ requires $R=0$.  The full Ricci scalar is
then
\begin{equation}
 R = g_{MN} R^{MN}
   = g_{\mu\nu} R^{\mu\nu} + g_{ij} R^{ij} + g_{i\mu} R^{i\mu}
   =g_{\mu\nu} R^{\mu\nu} = R^{(4)} \; ,
\end{equation}
using $g^{jk}R_{jk}=0$ along with the fact that $g_{\mu i}$ no longer
transforms under general coordinate transformations.

\item[--] \emph{compact/periodic}: the simplest compact space is a
  torus with periodic boundary conditions and a radius $r$ of the
  compactified dimension $y_i = y_i + 2\pi r$.
\end{itemize}
\bigskip

In addition to the Ricci scalar, the Einstein--Hilbert action contains
explicit dependency on the determinant of the metric
$\sqrt{-g^{(4+n)}}$.  Since the extra dimensions are flat and spatial,
the contribution to $\text{det}(g_{MN})\equiv g$ is at most a sign.
We assume that $\sqrt{-g}$ is synonymous with $\sqrt{|g|}$.  Now, it
is straightforward to simplify the higher--dimensional bulk action
\begin{alignat}{5}
  S_\text{bulk} &= - \frac{1}{2} \Mstar^{n+2} 
                   \int d^{4+n}z \; \sqrt{-g^{(4+n)}} \; R^{(4+n)} 
                   \notag \\
               &= - \frac{1}{2} \Mstar^{n+2} \; (2 \pi r)^n 
                    \int d^4 x \; \sqrt{-g^{(4)}} \; R^{(4)} 
                   \notag \\
               &\equiv - \frac{1}{2} \Mpl^2 \int d^4 x
                    \sqrt{-g^{(4)}} \; R^{(4)}.
\end{alignat}
In the last line we have matched the two theories, \ie we have assumed
that from a 4-dimensional point of view the actions have to be
identical, as long as we do not probe high enough energy scales to
observe quantum gravity effects.

This leads us to the basis of extra dimensions as a solution to the
hierarchy problem: our 4-dimensional Planck scale $\Mpl \sim
10^{19}$~GeV is not the fundamental scale of gravity. It is merely a
derived parameter which depends on the fundamental
$(4+n)$--dimensional Planck scale and the geometry of the extra
dimensions, \eg the compactification radius of the $n$--dimensional
torus. Matching the two theories translates into
\begin{equation}
\Mpl = \Mstar \; (2\pi r\Mstar)^{n/2}
\end{equation} 
If the proportionality factor $(2 \pi r \Mstar)^n$ is large we can
postulate that the fundamental Planck scale $\Mstar$ be not much
larger than $1\;\tev$. In that case the UV cutoff of our field theory
is of the same order as the Higgs mass and there is no problem with
the stability of the two scales.\bigskip

Assuming $\Mstar= 1\;\tev$ we can solve the equation above for the
compactification radius $r$ --- transferring the hierarchy problem
into space-time geometry:
\begin{center}
\begin{tabular}{c|c}
$n$ & $r$ \\
\hline
1 & $10^{12}$~m \\
2 & $10^{-3}$~m \\
3 & $10^{-8}$~m \\
... & ...       \\
6 & $10^{-11}$~m \\
\end{tabular}
\end{center}
At least in the simplest model $\delta=1$ is ruled out by classical
bounds on gravity as well as astrophysical data.  A possible exception
is if there is a non trivial mass gap between massless and massive
excitations~\cite{gps}.  For larger values of $n$ we need to test
Newtonian gravity at small distances~\cite{gravity_tests}. Note that
the analysis in this section is purely classical, and it is obvious
that its physical degrees of freedom do not survive
compactification. For this we resort to the original ideas of Kaluza
and Klein and decompose the higher-dimensional gravitational theory as
an effective 4-D theory with residual gauge
symmetries~\cite{kaluza_klein}.

%----------------------------------------------------------------------
\subsection{Gravitons in extra dimensions}

The first step towards a viable description of extra dimensional
effects in experiment is deriving the properties of spin-2 gravitons
in these extra dimensions~\cite{grw,coll_tao}.  Generically, a
massless graviton in higher dimensions can be described by an
effective theory of massive gravitons and gauge fields in four
dimensions.  The inclusion of massive spin two fields is particularly
interesting from a theoretical point of view since the Pauli--Fierz
mass term~\cite{fierz_pauli} and the coupling to matter fields is
highly restricted.  In particular, it is inconsistent to introduce
massive spin-2 fields not originating from some type of Kaluza--Klein
decomposition~\cite{kk_sugra}.\bigskip

We start with the $(4+n)$-dimensional Einstein equation
\begin{equation}
R_{AB}-\frac{1}{2} g_{AB} R \; = \; \frac{T_{AB}}{\Mstar^{2+n}}
\label{eq:ee}	
\end{equation}
and rewrite the metric in terms of our flat background metric
$\eta_{AB}$ and a fluctuating spin-2 field $h_{AB}$
\begin{equation}
 g_{AB} = \eta_{AB} + 2 \frac{h_{AB}}{\Mstar^{1+n/2}} \; .
\end{equation}
The prefactor ensures that $h$ (and with it the kinetic term in the
Lagrangian) has the appropriate mass dimension for a propagating
bosonic field $[h]=m^{(2+n)/2}$.  In terms of $h$, Einstein's equations
to linear order give
\begin{alignat}{5}
\Mstar^{1+n/2} \left( R_{AB}-\frac{1}{2} g_{AB} R \right) 
&=
  \Box h_{AB}
 -\p_A\p^C h_{CB} 
 -\p_B\p^C h_{CA}
 +\p_A\p_B h_C^C
 -\eta_{AB}\Box h^C_C
 +\eta_{AB}\p^C\p^D h_{CD} \notag \\
&= - \frac{T_{AB}}{\Mstar^{1+n/2}} \; .
\label{eq:EHactag}	
\end{alignat}
The equation of motion follows from the bilinear
action which we refer to as the linearized Einstein--Hilbert action:
\begin{alignat}{5}
\lag=&-\frac{1}{2}h^{MN}\Box h_{MN}+\frac{1}{2}h \Box h
-h^{MN}\p_M\p_N h +h^{MN}\p_M\p_Lh_L^N
-\Mstar^{-(1+n/2)}h^{MN}T_{MN},
\label{eq:KKLAG} 
\end{alignat}
The slightly circumvent logic (Einstein--Hilbert action$\rightarrow$
Einstein's equation $\rightarrow$ linearized Einstein's equation
$\rightarrow$ linearized Einstein--Hilbert action) leading us to
eq.(\ref{eq:KKLAG}) is necessary because the energy momentum tensor is
generated through
$T^{\mu\nu}= 2/\sqrt{-g} \; \delta S/\delta g_{\mu\nu}$
when computing the equations of motion.  Had we inserted the graviton
decomposition into the Einstein--Hilbert action directly, the resulting
linearized Einstein equations would describe a freely propagating
field.  The linearized variation analogous to eq.(\ref{eq:metvariation})
is
\begin{equation}
\delta h_{AB}=\p_A \xi^M +\p_B \xi^M \; ,
\label{eq:linvar}	
\end{equation}
leaving the linearized action invariant up to terms $\ope(\xi^2)$ with $h$ 
and $\xi$ treated as the same order.\bigskip

The ADD model breaks this full symmetry by compactifying the extra
dimensions.  Periodic boundary conditions allow us to Fourier
decompose the $y$ component of the graviton field
\begin{alignat}{5}
h_{AB}(z) 
&= \sum_{m_1=-\infty}^{\infty} \cdots
   \sum_{m_n=-\infty}^{\infty}
   \frac{h^{(\vec{n})}_{AB}(x)}{\sqrt{(2\pi r)^n}}  
	e^{i n_j y_j/r}
\notag
\\
&= h_{AB}^{(0)}(x)+ \sum_{n_1=1}^{\infty} \cdots
     \sum_{n_n=1}^{\infty}
     \frac{1}{\sqrt{(2\pi r)^n}} 
     \left[h^{(\vec{n})}_{AB}(x)e^{i n_j y_j/r}
    +h^{\dagger(\vec{n})}_{ AB}(x)e^{-i n_j y_j/r}\right]
\end{alignat}
where $h^{(\vec{n})}_{AB}(x)$ is a four dimensional bosonic field with mass
dimension one. The second step is possible because $h_{AB}(z)$ is 
real.  

To avoid confusion we emphasize that $h^{\dagger(\vec{n})}_{ AB}(x)$ does
not constitute an additional degree of freedom in the theory.  The
internal index $\vec{n}$ can be thought of as a discretized momentum index,
such that $h^{(\vec{n})}_{ AB}(x)$ and $h^{\dagger(\vec{n})}_{ AB}(x)$ differ only
by the sign of the extra-dimensional momentum $h^{\dagger(\vec{n})}_{
  AB}(x)=h^{(-\vec{n})}_{ AB}(x)$.  This is also obvious from the fact that
$h^{(\vec{n})}_{ AB}(x)$ and $h^{(\vec{n}')}_{ AB}(x)$ are not distinct field
excitations.  It is now simple to work out the form of Einstein's
equations in terms of the field $h^{(\vec{n})}_{AB}(x)$. For example, the
first term in eq.(\ref{eq:EHactag}) decomposes as
\begin{alignat}{6}
 \Box^{(4+n)} h_{AB}(z) &= \sum_{n_j} \frac{1}{(2\pi r)^{n/2}} \p_C\p^C
                \left[ h^{(\vec{n})}_{AB}(x) \; e^{i(n \cdot y)/r}
                \right]\notag\\
             &= \sum_{m_j} \frac{1}{(2\pi r)^{n/2}} \p_C
                \left[ 
                \left( \delta^C_\mu \p^\mu h^{(\vec{n})}_{AB}(x) 
                      + \delta^C_j h^{(\vec{n})}_{AB}(x) \frac{i n_j}{r} 
                \right) \; e^{i(n\cdot y)/r} 
                \right] \notag\\
             &= \sum_{m_j} \frac{1}{(2\pi r)^{n/2}}
                \left[ \Box^{(4)} 
                      - \frac{n^j n_j}{r^2}
                \right] \; h^{(\vec{n})}_{AB}(x) \; e^{i(n\cdot y)/r} \notag\\
%             &= \sum_{m_j} \frac{1}{(2\pi r)^{n/2}} \; e^{i(m\cdot y)/r}
%                \left(\Box^{(4)} +\hat{k}^2
%                \right) h^{(m)}_{AB}(x). 
\end{alignat}
Multiplying by $e^{-i(n\cdot y)/r}$ and using the energy momentum tensor
from eq.(\ref{eq:conEMT}) eliminates the exponential on the
right-hand side of eq.(\ref{eq:EHactag}).  An
independent check on the consistency of this method is that the
4-dimensional massive graviton field $h^{(\vec{n})}_{\mu\nu}(x)$ only has a
Pauli--Fierz mass term, $\propto \left[
h_{\mu\nu}-\eta_{\mu\nu}h\right]$ and no mass terms originating from
mixed index derivatives. This is required for a consistent spin-2
field~\cite{fierz_pauli}.\bigskip

From the Einstein equations we can brute force derive the action
bilinear in the fields $h^{(\vec{n})}_{AB}(x)$.  This field does not
transform irreducibly under the Lorentz group in four dimensions.  As
an ansatz we introduce a field decomposition~\cite{grw} which forms
irreducible representations.  Using the convenient definitions
$\hat{n}\equiv\vec{n}/r$ and $\kappa\equiv\sqrt{3(n-1)/(n+2)}$, the
action in terms of these new fields manifestly carries the correct
degrees of freedom:
\begin{alignat}{5}
G_{\mu\nu}^{(\vec{n})}
 &= h_{\mu\nu}
  +\frac{\kappa}{3} \left( \eta_{\mu\nu}
                          +\frac{\p_\mu \p_\nu}{\hat{n}^2} \right) H^{(\vec{n})}
  -\p_\mu\p_\nu P+\p_\mu Q_\nu+\p_\nu Q_\mu
\notag \\
V_{\mu j}^{(\vec{n})}
 &= \frac{1}{\sqrt{2}} \left( ih_{\mu j}-\p_\mu P_j-\hat{n}_{j} Q_\mu\right)
\notag \\
S_{jk}^{(\vec{n})}
 &= h_{jk} - \left( \eta_{jk} + \frac{\hat{n}_j\hat{n}_k}{\hat{n}^2} \right)
            \frac{H^{(\vec{n})}\kappa}{n-1}
    +\hat{n}_{j} P_k + \hat{n}_k P_j - \hat{n}_j\hat{n}_k P
\notag \\
H^{(\vec{n})}
 &= \frac{1}{\kappa}\left[h_j^j+\hat{n}^2P\right]
\notag \\
Q_\mu^{(\vec{n})}
 &= -i\frac{\hat{n}_j}{\hat{n}^2}h^j_\mu
\notag \\
P_{j}^{(\vec{n})}
 &= \frac{\hat{n}_{k}}{\hat{n}^2}h^k_j + \hat{n}_jP
\notag \\
P^{(\vec{n})}
 &= \frac{\hat{n}_k \hat{n}_j}{\hat{n}^4}h_{jk}
\label{eq:newfields}
\end{alignat}
The fields $Q_\mu, P_{j}$ and $P$ are not invariant under general
coordinate transformations eq.(\ref{eq:linvar}) and cannot appear
independently in the effective 4-dimensional action.  In this sense
they are gauge degrees of freedom and setting $Q_\mu=P_{j} =P=0$
corresponds to a unitary gauge with no propagating ghosts.  

The decompositions is similar in spirit to the well-known
Kaluza--Klein~\cite{kaluza_klein} decomposition where a 5-dimensional
metric $g_{AB}$ is decomposed into a 4-dimensional metric
$g_{\mu\nu}$, a vector $A_\mu$ and a scalar $\phi$.  At the massless
level these fields decouple and the five degrees of freedom for a
5-dimensional graviton decompose appropriately as $2+2+1$.  Including
masses the vector and scalar fields are eaten by the graviton to build
a massive 4-dimensional graviton with five degrees of freedom.

Similarly, starting from eq.(\ref{eq:newfields}) our unitary gauge
choice allows $G_{\mu\nu}^{\vec{n}}$ to become massive by eating $P$ and
$Q_\mu$.  However, as opposed to the 5-dimensional case this does not
exhaust the degrees of freedom; there is an additional massive
$(n-1)$-multiplet of vectors $V_{\mu j}$ which eat $P_j$ to obtain their
longitudinal polarization.  Finally, there are $(n^2-n-2)/2$ scalars
in the symmetric tensor $S_{jk}$ as well as the singlet scalar
$H$. The total number of degrees of freedom is
\begin{equation} 
1\cdot5
\; + \; (n-1)\cdot 3
\; + \;  \frac{n^2-n-2}{2} \cdot 1
\; + \;  1 \cdot 1
=\frac{(n+4)(n+1)}{2} \; .
\label{eq:numberofcons} 	
\end{equation}
A similar analysis of $(4+n)$-dimensional gravity gives an identical
counting of degrees of freedom: the only physical field is a symmetric
tensor $h_{AB}$ with $(4+n)(5+n)/2$ components. These are reduced by
fixing the gauge; typically, the harmonic condition $\p_A h^A_B=\p
h^A_A/2$ amounts to $4+n$ constraints.  Furthermore, we are free to
add terms to the variation parameter in eq.(\ref{eq:linvar}) with
$\Box\xi_M$ leaving the action invariant.  Altogether there are
$2(4+n)$ constraints, and counting of degrees of freedom is identical
to eq.(\ref{eq:numberofcons}).\bigskip

In terms of the new physical fields Einstein's equations simplify to
\begin{alignat}{5}
(\Box+\hat{n}^2)& G_{\mu\nu}^{(\vec{n})}
&&= \frac{1}{\Mpl} \left[ -T_{\mu\nu}
                         +\left(\frac{\p_\mu \p_\nu}{\hat{m}^2}
                               +\eta_{\mu\nu}\right)
                          \frac{T_\lambda ^\lambda}{3}
                  \right] \notag \\ 	
(\Box+\hat{n}^2)& V_{\mu j}^{(\vec{n})} &&= 0 \notag \\
(\Box+\hat{n}^2)& S_{jk}^{(\vec{n})}    &&= 0 \notag \\
(\Box+\hat{n}^2)& H^{(\vec{n})}         &&= \frac{\kappa}{3\Mpl} T_\mu^\mu
\label{eq:ee_new}
\end{alignat}
so that the linearized Lagrangian eq.(\ref{eq:KKLAG}) in terms of the
fields eq.(\ref{eq:newfields}) reads (omitting the sum over the index
$\vec{n}$ for all fields)
\begin{alignat}{5}
\lag \sim & -\frac{1}{2}G^{\dagger \mu\nu}(\Box+\mkk^2)G_{\mu\nu}
                        -\frac{1}{2}G^{\dagger \mu}_\mu (\Box+\mkk^2)G^\nu_\nu
                        -G^{\dagger \mu\nu} \p_\mu\p_\nu G^{\lambda}_{\lambda}
\notag \\
&+ G^{\dagger\mu\nu} \p_\mu\p_{\lambda}G^{ \lambda}_\mu
 -\frac{1}{2}H^{ \dagger}(\Box+\mkk^2)H
\notag \\
&-\frac{1}{\Mpl} \left[G^{\mu\nu}
                       -\frac{\kappa}{3}\eta^{\mu\nu} H
                 \right]T_{\mu\nu} + \cdots
\label{eq:effectiveaction}
\end{alignat}
where the ellipses stand for free field kinetic terms.  Here and
henceforth we define $\mkk^2\equiv\hat{n}^2=\vec{n}^2/r^2$.  The
structure of Einstein's equations eq.(\ref{eq:ee_new}) reveals a few
particularities: the fields $V_{\mu j}$ and $S_{jk}$ do not couple to
the energy momentum tensor, \ie to the Standard Model.  The massive
gravitons $G_{\mu\nu}$ do couple to the Standard Model. Their Fourier
coordinate only appears as a mass-squared $\hat{n}^2$ and in the
coupling to the trace of the energy-momentum tensor. This means their
couplings are level--degenerate and their masses and couplings depend
only on the length, but not on the orientation of the vector
$\hat{n}$.\bigskip

We focus on the properties of conformally invariant theories, where
$T^\mu_\mu=0$, because this is a good approximation of all relevant
particle masses as compared to the LHC energy. For such massless
theories 
\begin{equation}
 (\Box+\mkk^2) \; G_{\mu\nu}^{(m)}=
        - \frac{T^{\mu\nu}}{\Mpl} 
\end{equation}
describes physical gravitons produced by quark or gluon interactions
and either vanishing or decaying to leptons.  The scalar mode
$H$ plays a special role.  Its massless radion mode corresponds
to a fluctuation of the volume of the compactified extra dimension. We
assume that the compactification radius $r$ is stabilized in some
way~\cite{stabilize}, giving mass to the radion~\cite{radion}.  More
importantly, the radion only couples to a massive theory, so it is not
surprising that as a scalar with no Standard Model charge it will mix
with a Higgs boson without too drastic effects.\bigskip

Before deriving Feynman rules we will briefly outline the significance
of Kaluza--Klein towers of massive gravitons: first of all, the basic
relation derived at the classical level can be rewritten as $\Mpl^2
\equiv \Mstar^2 N$ where $N \sim (2\pi r\Mstar)^n$ is the number of
Kaluza--Klein species existing below the scale $\Mstar$.  A heuristic
argument for this relies on the spacing between consecutive KK modes
$\delta \mkk \sim 1/r$, so that $r \Mstar= \Mstar/\delta \mkk$ gives
the number of KK modes with the vector $\vec{n}$ occupied only in one
direction \ie $\vec{n}=(j,0,0,\cdots)$ with $j$ being some integer.  A
generic vector has $n$ such directions, so in general there are $(r
\Mstar)^{n}$ possibilities.  Realistic numbers for $\Mstar \sim$~TeV
give $N\sim 10^{32}$.  A similar result is achieved by considering
black hole evaporation~\cite{gia_bh}. This multiplicity of states is
what determines the visible effects of ADD models at the LHC.\bigskip

As a side remark, it is not altogether mysterious that we are summing
over a very large number of KK states.  In $(4+n)$-dimensional
language the graviton propagator is simply $1/(p^A p_A)$.  The
momentum $p^A$ obeys momentum conservation at each vertex to two
Standard Model particles.  This way 4-dimensional external lines fix
the momentum in four directions, leaving an integration over $p_j$
\begin{equation}
\int d^n p_j \frac{1}{p^A p_A}
= \int d^n p_j \frac{1}{p^\mu p_\mu-p^j p_j} 
\sim \sum_{\mkk} \frac{(\delta \mkk)^n}{p^\mu p_\mu- \mkk^2} \; .
\end{equation}
For KK modes as intermediate states, proper treatment of the 
KK tower implies a closed integral,
similar to an additional $n$-dimensional loop integral $\int\mat$.  A
similar argument reveals that the additional momentum directions
available for final state KK particles amounts to a modified phase
space integral $\int |\mat|^2$.\bigskip

To summarize our main results relevant for ADD phenomenology; the
spin-2 graviton field couples to the energy momentum tensor
universally for massless states, suppressed by the 4-dimensional
Planck scale $\Mpl$. This predicts the production of massive gravitons
at the LHC from gluon as well as quark initial states. The index
structure of the massive graviton propagator and vertices are a mess,
but theoretically well defined in our effective field theory. There are
a large number of gravitons organized by a KK tower which again couple
universally to Standard Model fields. The mass splitting between the
KK states inside the tower is given by $1/r$ which translates into
($\Mstar=1$~TeV as before):
\begin{alignat}{5}
\delta
\mkk\sim\frac{1}{r}=2\pi\Mstar\left(\frac{\Mstar}{\Mpl}\right)^{2/n}&=
\left\{
\begin{array}{lll}
\displaystyle{0.003~\text{eV}}
\qquad &&(n=2) \\
\displaystyle{0.1~\text{MeV}}
\qquad &&(n=4) \\
\displaystyle{0.05~\text{GeV}}
\qquad &&(n=6)
\end{array} \right.
\end{alignat}
On the scale of high-energy experiments or the weak scale ($m_Z \sim
91$~GeV), this mass splitting is tiny.  For example the LHC will be
unable to resolve such mass differences, which allows us to generally
replace the sum over graviton modes (either as intermediate states or
as final states) by an integration over a continuous variable.  We
will show this conversion to an integral in the next section. What we
can also see from this mass splitting is that gravitons in this model
might well be stable, just because they are too light to decay to two
Standard Model particles even via a gravitational interaction. A KK
tower of gravitons appears as missing energy at the LHC.

%----------------------------------------------------------------------
\subsection{Feynman rules}

To lowest order in the graviton field the coupling to massless matter
is given by eq.(\ref{eq:effectiveaction}).  We illustrate the
extraction of the vertices from the energy-momentum tensor in a
manifestly symmetric way using the QED Lagrangian
\begin{equation}
\lag_\text{QED} = 
\frac{\sqrt{-g}}{\bar{M}_P}
\left( i\bar{\psi}\gamma^a \mathcal{D}_a \psi
      -\frac{1}{4} F_{\mu\nu}F^{\mu\nu}
\right)
\end{equation}
where the covariant derivative contains a gauge and coordinate
connection.  Taking the variation in the metric for only the gauge
field and noting $\delta\sqrt{-g}=-\sqrt{-g} \; g^{\mu\nu} \; \delta
g_{\mu\nu}/2$ we find
\begin{alignat}{5} 
T^{\mu\nu}_\text{gauge} 
&=\frac{2}{\sqrt{-g}} \; \frac{\delta}{\delta g_{\mu\nu}} \; \sqrt{-g}\; 
  \left( -\frac{1}{4} F_{\rho\sigma}F_{\alpha\beta} \; g ^{\rho\alpha} g^{\sigma\beta}
  \right) \notag \\
&=\frac{\eta^{\mu\nu}}{4} F_{\alpha\beta}F^{\alpha\beta}
  +F^\mu_\alpha F^{\alpha \nu}
\end{alignat}
and for the the purely fermionic contribution 
\begin{equation}
T^{\mu\nu}_\text{fermion} 
= \frac{i}{4}\bar{\psi} 
  \left( \p^\mu \gamma^\nu
        +\p^\nu \gamma^\mu \right) \psi 
        -\frac{i}{4} \left( \p^\mu\bar{\psi} \gamma^\nu
                           +\p^\nu\bar{\psi} \gamma^\mu \right) \psi
\label{eq:Tmn} 
\end{equation}
All momenta are incoming to the vertex.  To derive the Feynman rules
we need to symmetrize the graviton and gauge boson indices
separately.\bigskip

In the following, we will quote the Feynman rules relevant to our LHC
analysis. Because gravity couples to every particle in and beyond the
Standard Model there are in fact many other graviton
vertices~\cite{grw,coll_tao}. The fermion--graviton and
gluon--graviton vertices are
\bigskip
\begin{equation*} 
\parbox{30mm}{
\begin{fmfgraph*}(35,55)
   \fmfleft{w,we}\fmfright{g}
   \fmf{fermion}{w,v,we}\fmf{double}{v,g}
   \fmflabel{$f(k_1)$}{w}\fmflabel{$\bar{f(k_2)}$}{we}
   \fmflabel{$G_{\mu\nu}$}{g}
   \fmfdot{v}
\end{fmfgraph*}}
=
-\frac{i}{4\Mpl}\left[W_{\mu\nu}+W_{\nu\mu}\right] 
\end{equation*} \smallskip
and \smallskip
\bigskip
\begin{equation*}
\parbox{30mm}{
 \begin{fmfgraph*}(35,55)
   \fmfleft{w,we}\fmfright{g}
   \fmf{boson}{w,v,we}\fmf{double}{v,g}
   \fmflabel{$\epsilon^b_{\alpha}(k1)$}{w}\fmflabel{$\epsilon^a_{\beta}(k_2)$}{we}
   \fmflabel{$G_{\mu\nu}$}{g}
   \fmfdot{v}
\end{fmfgraph*}}
=
-\frac{i\delta_{ab}}{\Mpl}\left[W_{\mu\nu\alpha\beta}+W_{\nu\mu\alpha\beta}\right]
\end{equation*}
with
\begin{alignat}{5}
W_{\mu\nu} =& (k_1-k_2)_\mu\gamma_\nu
 \notag \\
W_{\mu\nu\alpha\beta} =&
\frac{1}{2} \eta_{\mu\nu} \left( k_{1\beta} k_{2\alpha} 
                               -k_1\cdot k_2 \eta{\alpha\beta} \right)
+\eta_{\alpha\beta}k_{1\mu}k_{2\nu}
+\eta_{\mu\alpha} \left( k_1\cdot k_2\eta_{\nu\beta}
                       -k_{1\beta}k_{2\nu} \right)
-\eta{\mu\beta}k_{1\nu}k_{2\alpha}
\end{alignat}
The non-abelian part of the field strength does not contribute, and
the color factor $\delta_{ab}$ reflects the fact that the graviton is
a gauge singlet.  The final ingredient needed is the graviton
propagator, which is the momentum-space inverse two-point function
from the bilinear terms in eq.(\ref{eq:effectiveaction}).  This
Lagrangian describes the physical fields and --- as opposed to the
massless graviton --- no additional gauge fixing is required.
\begin{equation*}
\parbox{23mm}{
\begin{fmfgraph*}(35,55)
   \fmfleft{w}\fmfright{g}
   \fmf{double}{w,g}
   \fmflabel{$G_{\mu\nu}(k)$}{w}
   \fmflabel{$G_{\alpha\beta}^{\dagger}(k)$}{g}
\end{fmfgraph*}}
\quad =\frac{iP_{\mu\nu\alpha\beta}}{k^2-m^2}
\end{equation*}
with
\begin{alignat}{5}
 P_{\mu\nu\alpha\beta}=\frac{1}{2}&(\eta_{\mu\alpha}\eta_{\nu\beta}+
\eta_{\mu\beta}\eta_{\nu\alpha}-\eta_{\mu\nu}\eta_{\alpha\beta})
\notag
\\
&-\frac{1}{2\mkk^2}(\eta_{\mu\alpha}k_\nu k_\beta + \eta_{\nu\beta} k_\nu 
k_\alpha + \eta_{\mu\beta} k_\nu k_{\alpha} + \eta_{\nu\alpha}k_\mu k_\beta)
\notag
\\
&+ \frac{1}{6}\left(\eta_{\mu\nu}+\frac{2}{\mkk^2}k_\mu k_\nu\right)
\left(\eta_{\alpha\beta}+\frac{2}{\mkk^2}k_\alpha k_\beta\right).
\label{eq:Propo} 
\end{alignat}
It is easy to recognize the first line in eq.(\ref{eq:Propo}) as
(ignoring overall normalization) the massless graviton in the De
Donder gauge.  A good exposition on different forms (different weak
field expansions) of the massive and massless propagator is given in
Ref.\cite{propagator}.\bigskip

The amplitude for a generic $s$-channel process mediated by virtual gravitons
will then look like
\begin{alignat}{5}
\amp&\sim 
              \frac{1}{\Mpl^2}
              \sum T_{\mu\nu} \;
                   \frac{P_{\mu\nu\alpha\beta}}{s-\mkk^2} \;
                   T_{\alpha\beta}
                   \notag\\
           &= \frac{1}{\Mpl^2}
              \sum T_{\mu\nu} \;
                          \frac{ \eta_{\mu\alpha}\eta_{\nu\beta}
                                +\eta_{\mu\beta} \eta_{\nu\alpha}
                                -\eta_{\mu\nu}   \eta_{\alpha\beta}
                                +\eta_{\mu\nu}\eta_{\alpha\beta}/3}
                               {2(s-\mkk^2)} \;
                   T_{\alpha\beta}
                   \notag\\
           &= \frac{1}{\Mpl^2}
              \sum \frac{1}{s-\mkk^2} \; T_{\mu\nu} T^{\mu\nu} \notag \\
           &\equiv \s(s) \; \mathcal{T} \; .
\label{eq:d8}
\end{alignat}
On the way we use the conservation and tracelessness of the energy
momentum tensor \ie $T^{\mu}_{\mu}=k_{\mu} T^{\mu\nu} =0$.  This form
is useful because the field content $\mathcal{T}\equiv T_{\mu\nu}
T^{\mu\nu}$ and an appropriate coefficient form a general dimension-8
operator $\s \mathcal{T}$.\bigskip

In addition, a loop-induced dimension-6 operator will be generated by
diagrams of the form
\begin{equation*}
\parbox{30mm}{
    \begin{fmfgraph*}(50,30)
 	\fmfbottom{i1,d1,o1}
	\fmftop{i2,d2,o2}
	\fmf{fermion}{i1,v1,v2,o1}
	\fmf{fermion}{o2,v4,v3,i2}
	\fmf{double,tension=0}{v1,v3}
	\fmf{photon,tension=0}{v2,v4}
    \end{fmfgraph*}}
\quad \quad
\parbox{30mm}{
    \begin{fmfgraph*}(50,30)
 	\fmfbottom{i1,d1,o1}
	\fmftop{i2,d2,o2}
	\fmf{fermion}{i1,v1,v2,o1}
	\fmf{fermion}{o2,v4,v3,i2}
	\fmf{double,tension=0}{v1,v3}
	\fmf{double,tension=0}{v2,v4}
    \end{fmfgraph*}}
\end{equation*}
The resulting four-fermion interactions couples two axial-vector
currents
\begin{equation}
c_6 \sum \left(\bar{\psi} \gamma_5 \gamma_\mu \psi \right)^2
\end{equation}
where the sum is over all fermions in the theory.  The coefficient
$c_6$ can be estimated by naive dimensional analysis~\cite{gs}.  Such
KK graviton contributions can be compared to effects from a modified
theory with non-symmetric connection~\cite{non_symmetric_conn}.  Generically,
all these loop effects will affect electroweak precision
observables.\bigskip

Our final consideration is real graviton emission off any kind of
Standard Model process, preferably a so-called standard
candle~\cite{sm_candles} which we expect to understand well from a theory
as well as an experimental perspective. Calculating amplitudes
corresponding to diagrams such as
\begin{equation*}
\parbox{25mm}{
\begin{fmfgraph*}(85,55)
  \fmfleft{em,ep}
  \fmf{gluon}{em,Zee,ep}
  \fmf{gluon}{Zee,Zff}
  \fmf{gluon}{fb,Zff}
  \fmf{double}{Zff,f}
  \fmfright{fb,f} 
  \fmfdot{Zee,Zff}
\end{fmfgraph*}}
\end{equation*}
requires polarization tensors for external gravitons. We can for
example construct five tensors $\epsilon_{\mu\nu}^{s}$ where
$s=(1,2,\cdots 5)$ by taking outer products of the three massive gauge
boson polarization vectors.  A convenient parameterization is given in
Ref.~\cite{coll_tao}.  Most importantly, the $\epsilon_{\mu\nu}^{s}$
obtained this way satisfy
\begin{equation}
\sum_{s} \epsilon_{\mu\nu}^s(k) \epsilon_{\alpha\beta}^s(k)=P_{\mu\nu\alpha\beta}(k)
\label{eq:polsum}
\end{equation}
when summed over all polarization states. This brief review of
calculational details now puts us into a position to discuss LHC
processes.

%----------------------------------------------------------------------
\subsection{Collider observables}

In this section we summarize possible direct and indirect signatures
for massive Kaluza--Klein gravitons at
colliders~\cite{grw,coll_joanne,coll_maxim,more_and_more,review_joanne}.
There is also a large amount of phenomenological work confronting
electroweak precision data~\cite{tao_danny} or astrophysical
data~\cite{add_astro} and large extra dimensions, in part orthogonal
to their collider effects~\cite{gps}, which we will not have space to
cover here. Current limits strongly constrain ADD models with few
extra dimensions favoring $n>2$.  As we will see in the following
sections, such a scenario is also the most conceptually
interesting. For two to seven extra dimensions, strong direct
constraints on $\Mstar$ come from recent Tevatron
data~\cite{tevatron_real,tevatron_virt}.\bigskip

Of the two classes of collider observables we first consider the real
emission of Kaluza--Klein gravitons at the LHC~\cite{grw,lhc_real}.
The outgoing gravitons cannot be detected in our detectors --- similar
to neutrinos or possible dark matter agents --- so they appear as
missing transverse momentum or missing transverse energy
$\slashed{E}_T$. One process to radiate gravitons off is single jet
production~\cite{coll_maxim}.  The Feynman rules discussed above allow
us to compute squared-averaged amplitudes for partonic sub-processes
such as $qg\rightarrow qG$, $q\bar{q}\rightarrow gG$ and $gg
\rightarrow gG$ all of which lead to the same final state: one hard
QCD jet and missing transverse energy $\slashed{E}_T$. Due to the
strong QCD coupling this is the most likely real emission search
channel.

For the graviton--jet final state there is an obvious irreducible
background coming from $q\bar{q}\rightarrow Zg$ where the gluon is
emitted from an initial-state quark and the $Z$ decays into neutrinos.
This background is known to next-to-leading order~\cite{zj_nlo}, but
at large partonic event energies the theoretical rate prediction
becomes increasingly hard, due to large logarithms. Extracting new
physics from pure QCD signatures at the LHC will therefore always be
tough and somewhat dangerous (as we have seen in the past at the
Tevatron, where many signals for new physics have come and gone over
the years).\bigskip

Due to the structure of the parton densities of quarks and gluons
inside a proton, Tevatron searches for large extra dimensions
concentrate on $\gamma\slashed{E}_T$ final states. Similarly, at the
LHC a one photon final state could be resolved in the detectors
optimized for Higgs searches in the $H \to \gamma \gamma$ decay
channel. Hard single photon events would constitute a revealing
signature for physics beyond the Standard Model.

Similarly, the Drell--Yan process $q \bar{q} \to \gamma^*,Z \to \ell^+
\ell^-$ with two leptons (electrons or muons) in the final state is
the arguably best known hadron collider process~\cite{dy}. A large
amount of missing energy in this channel would be a particularly clean
signal for physics beyond the Standard Model at the LHC~\cite{dave}.
Depending on the detailed analysis, both of these electroweak
signatures do have smaller rates than a jet+graviton final state, but
the lack of QCD backgrounds and QCD-sized experimental and theory
uncertainties result in discovery regions of similar
size~\cite{grw,lhc_real}.\bigskip

Going back to the theoretical basis, the partonic cross section for
the emission of one graviton is not the appropriate observable. What
we are interested in is the entire KK tower contributing to the missing
energy signature
\begin{alignat}{5}
d\sigma^\text{tower}&=\sum_{\vec{n}}  d\sigma^\text{graviton}
=\int dN \; d\sigma^\text{graviton}
\end{alignat} 
where $\int dN$ is an integration over an $n$-dimensional sphere in KK
density space
\begin{equation}
\int dN \equiv  S_{n-1} \; |\vec{n}|^{n-1} \; d|\vec{n}|
\qquad \qquad
S_{n-1}=\frac{2\pi^{n/2}}{\Gamma(n/2)}
\end{equation}
In the ultraviolet the sum over $\vec{n}$ is truncated to those states
which satisfy kinematic constraints.  In particular, the KK mass
satisfies $\mkk = |\vec{n}|/r<\sqrt{s}$ where $\sqrt{s}$ is the partonic
center of mass energy (related to the proton center of mass energy via
$s = (\text{14 TeV})^2\; x_1 x_2$).

The KK state density we can rewrite into a mass density kernel using
$d\mkk/d|\vec{n}|=1/r$
\begin{equation}
dN=S_{n-1} \, r^n \, \mkk^{n-1} \, d \mkk 
  = \frac{S_{n-1}}{(2\pi\Mstar)^n} \left( \frac{\Mpl}{\Mstar}
                                       \right)^2 \, \mkk^{n-1} \, d\mkk \; .
\end{equation}
This implies for the production of a Kaluza--Klein tower
\begin{alignat}{2}
%d&\sigma^\text{one \; graviton}&&= |\mat|^2 \,
%                               (2\pi)^4 \, \delta^4(p_i-p_f)
%                               \frac{d\Phi_f}{F(p_1,p_2)} \\
d \sigma^\text{tower}  
= d\sigma^\text{graviton} \;
  \frac{S_{n-1} \, \mkk^{n-1} \, d \mkk}{(2\pi\Mstar)^n}
  \; \left( \frac{\Mpl}{\Mstar} \right)^2 \; .
\label{eq:KKTxsec}
\end{alignat}
The key aspects of this formula are:
\begin{itemize}
\item[--] The factor $M^2_\text{Planck}$ from the KK tower summation
  can be absorbed into the one-graviton matrix element squared. The
  effective coupling of the entire tower at the LHC energy scale $E$
  is then $E/\Mstar \gtrsim 1/10$ instead of $E/\Mpl$, \ie roughly of
  the same size as the Standard Model gauge couplings.
\item[--] In particular for larger $n$ the integral is infrared finite
  with the largest contributions arising from higher mass modes.  This
  is the effect that KK modes are more tightly spaced as we move to
  higher masses and even more so for a increasing number of extra
  dimensions $n$.
\item[--] Although $\mkk$ appears explicitly in the polarization sum
  and thus is naively present in the amplitude, it does not appear
  once we square the amplitude due to the arguments following
  eq.(\ref{eq:d8}).  The $\mkk$ integration at least on the
  partonic level --- \ie without the parton densities --- can be done
  without specifying the process.
\begin{alignat}{5}
\int dN =
\int_0^{\cutoff} \, \frac{S_{n-1}}{(2\pi\Mstar)^n}
                           \left( \frac{\Mpl}{\Mstar}
                           \right)^2 \, \mkk^{n-1} \, d \mkk  
%             &= S_{n-1} \, \frac{1}{(2\pi\Mstar)^n} 
%                           \left( \frac{\Mpl}{\Mstar}
%                           \right)^2 \, \frac{m^n}{n} \Bigg|^\mu_0 \notag \\
             &= \frac{S_{n-1}}{(2\pi\Mstar)^n}
                           \left( \frac{\Mpl}{\Mstar}
                           \right)^2 \, \frac{\cutoff^n}{n}
\label{eq:div_real}
\end{alignat}
  In this form we indeed see that our effective theory of KK gravity
  requires a cutoff to regularize an ultraviolet divergence, simply
  reflecting the fact that gravity is not perturbatively
  renormalizable. The crucial question becomes if the prediction of
  LHC observables is sensitive to $\cutoff$.
\item[--] For real graviton production the kinematic constraint
  $\mat=0$ for $\mkk > \sqrt{s}$ provides a natural ultraviolet cutoff
  on the $n$-sphere integration.  Therefore the result is insensitive
  to physics far above the LHC energy scale, which might or might not
  cover the fundamental Planck scale.
\end{itemize}
\bigskip

%----------------------------------------
\begin{figure}[t]
\includegraphics[width=8cm]{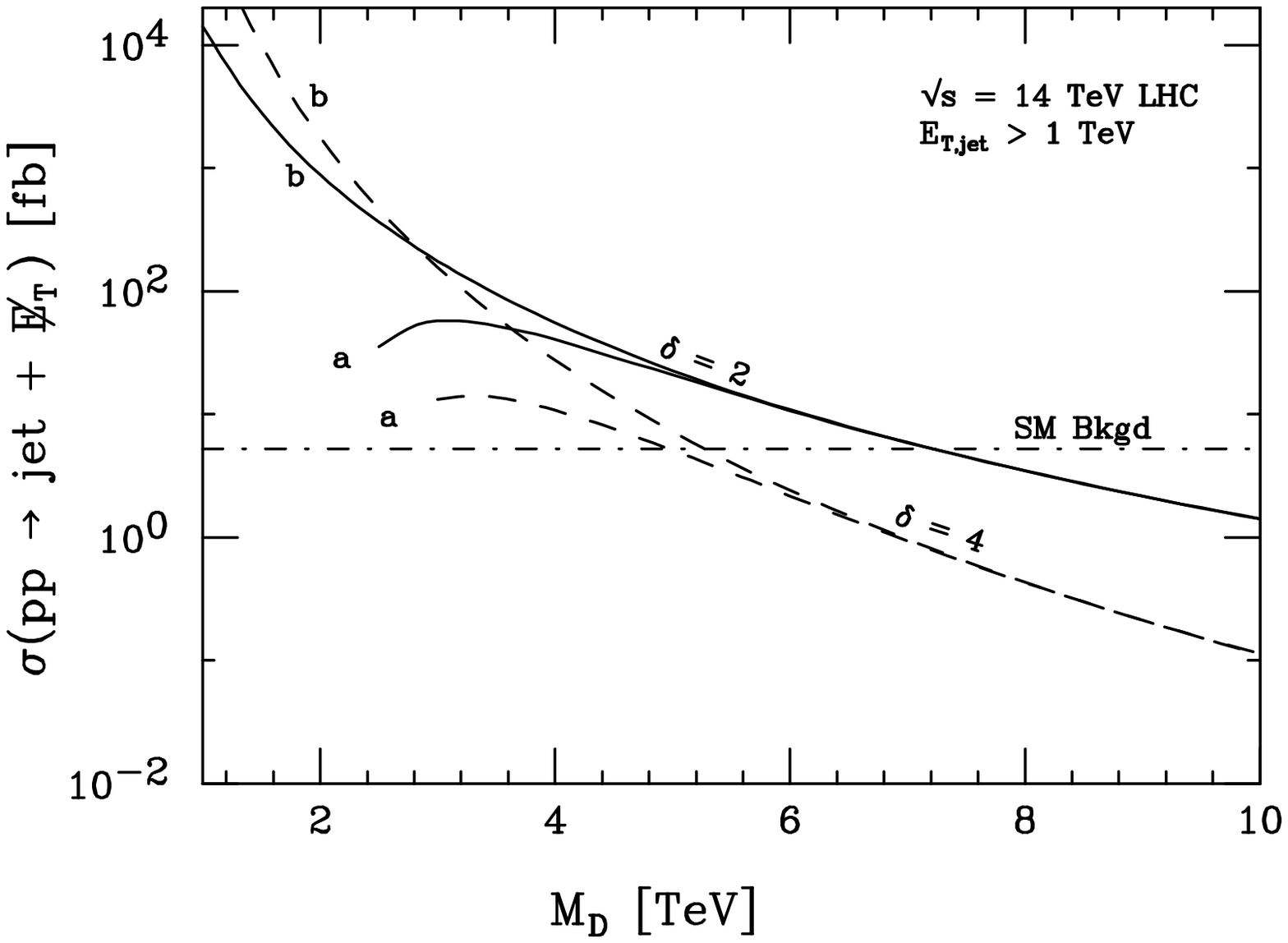}
\hspace*{5mm}
\raisebox{4mm}{\includegraphics[width=5.6cm]{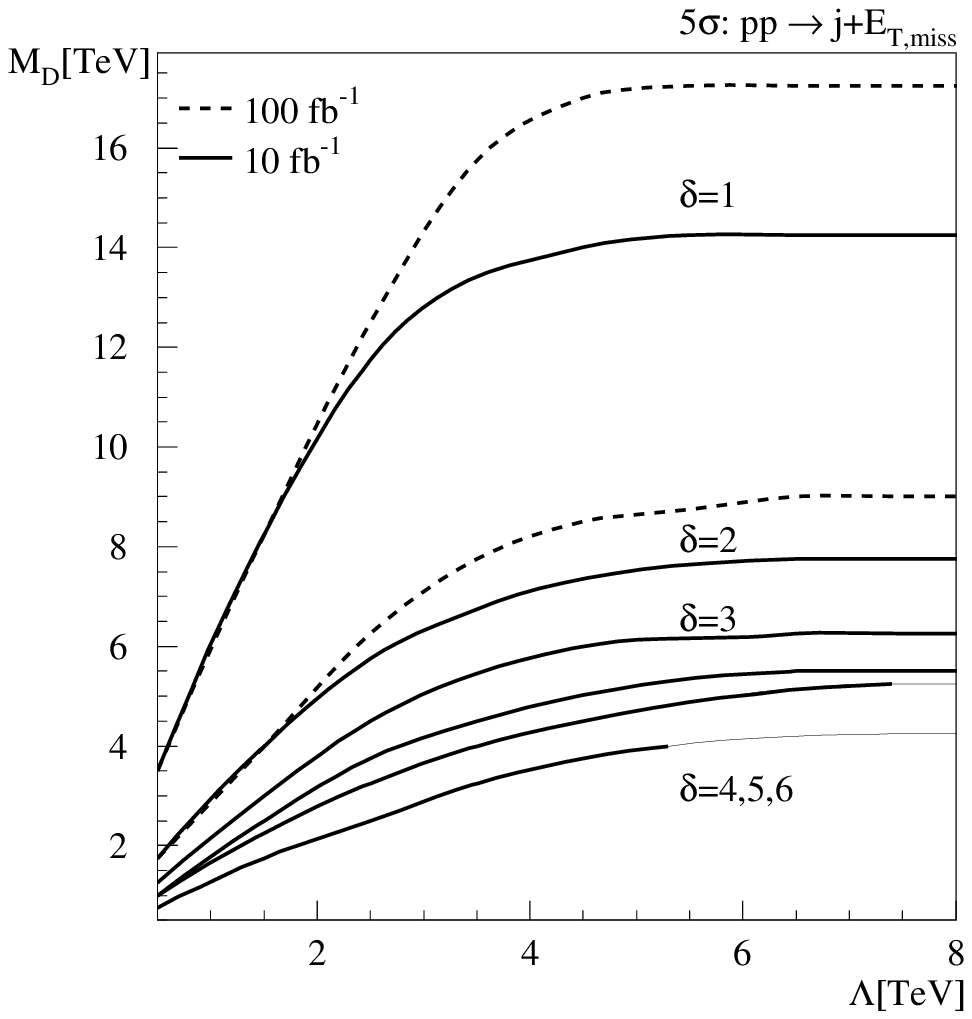}}
\caption{Left: production rates for graviton--jet production at the
  LHC.  The cut on the transverse mass of the jet indirectly acts as a
  cut on the transverse momentum from the graviton tower, without the
  additional experimental smearing from measuring missing transverse
  energy.  For the curve (a) a cutoff procedure sets $\sigma(s)=0$
  whenever $\sqrt{s}>\Mstar$. For curve (b) the $\mkk$ integration in
  eq.(\protect\ref{eq:KKTxsec}) includes the region $\sqrt{s}>\Mstar$. Figure
  from Ref.~\cite{grw}. Right: $5 \sigma$ discovery contours for real
  emission of KK gravitons in the plane of $\Mstar$ and the UV cutoff
  on $\sigma(s)$.  The transition to thin lines indicates a cutoff
  $\cutoff$ above $\Mstar$.  Figure from Ref.~\cite{gps}.  Note
  that $M_D$ in both figures corresponds to $\Mstar$ in the text.}
\label{fig:ReEmm}
\end{figure}
%----------------------------------------

In the left panel of Fig.\ref{fig:ReEmm} we see that the jet +
$\slashed{E}_{T}$ cross section becomes seriously dependent on physics
above $\Mstar$ the moment the Planck scale enters the range of
available energies at the LHC $\sqrt{s} \lesssim 3$~TeV. Above this
threshold the difference between curves (a) and (b) is small.

In the region where the curves differ significantly, UV effects of our
modelling of the KK spectrum become dominant and any analysis based on
the Kaluza--Klein effective theory will fail. Luckily, the parton
distributions, in particular the gluon density, drop rapidly towards
larger parton energies.  This effects effectively constrains the
impact of the UV region which includes $\sqrt{s}>\Mstar$.

In the right panel of Fig.\ref{fig:ReEmm} we see the $5\sigma$
discovery reach at the LHC with a variable ultraviolet cutoff
$\cutoff$ on the partonic collider energy. For each of the lines there
are two distinct regimes: for $\cutoff < \Mstar$ the reach in $\Mstar$
increases with the cutoff. Once the cutoff crosses a universal
threshold around 4~TeV the discovery contours reach a plateau and
become cutoff independent. This universal feature demonstrates and
quantifies the effect of the rapidly falling parton densities.  The
fact that the signal decreases with increasing dimension shows that
the additional volume element from the $n$-sphere integration is less
than the $1/\Mstar$ suppression from each additional
dimension. \bigskip

Virtual gravitons at the LHC demand a markedly different analysis
since by definition these signals do not produce gravitational missing
energy. The dimension-8 operator $\s(s) \mathcal{T}$ as shown in
eq.(\ref{eq:d8}) is induced by integrating out a whole graviton tower
exchange in $s$-channel processes at the LHC such as
\begin{equation*}
\parbox{25mm}{
\begin{fmfgraph*}(85,55)
  \fmfleft{em,ep}
  \fmf{gluon}{em,Zee,ep}
  \fmf{double}{Zee,Zff}
  \fmf{fermion}{fb,Zff,f}
  \fmfright{fb,f} 
  \fmfdot{Zee,Zff}
\end{fmfgraph*}}
\qquad \qquad \qquad \qquad
\parbox{25mm}{
\begin{fmfgraph*}(85,55)
  \fmfleft{em,ep}
  \fmf{gluon}{em,Zee,ep}
  \fmf{double}{Zee,Zff}
  \fmf{gluon}{fb,Zff,f}
  \fmfright{fb,f} 
  \fmfdot{Zee,Zff}
\end{fmfgraph*}}
\qquad \qquad \qquad \qquad
\parbox{25mm}{
\begin{fmfgraph*}(85,55)
  \fmfleft{em,ep}
  \fmf{gluon}{em,Zee,ep}
  \fmf{double}{Zee,Zff}
  \fmf{photon}{fb,Zff,f}
  \fmfright{fb,f} 
  \fmfdot{Zee,Zff}
\end{fmfgraph*}}
\end{equation*}

In the Standard Model some of these final states, leptons and weak
gauge bosons, can only be produced by a $q\bar{q}$ initial
state. Because at LHC energies the protons mostly consist of gluons,
such indirect graviton signatures get a head start.  The Tevatron
mostly looks for in two-photon or two-electron final
states~\cite{tevatron_virt}.  At the LHC the cleanest signal taking
into account backgrounds as well as experimental complications is a
pair of muons~\cite{coll_joanne}. In Higgs physics the corresponding
channel $H\rightarrow ZZ \rightarrow 4\mu$ is referred to as the
`golden channel', because it is so easy to extract.\bigskip

In the Standard Model the Drell--Yan process mediates muon pair
production via the $s$-channel exchange of on-shell and off-shell
$\gamma$ and $Z$ bosons. Aside from the squared amplitude for graviton
production, these Standard Model amplitudes interfere with the
graviton amplitude, affecting the total rate as well as kinematic
distributions. This mix of squared amplitudes and interference effects
make it hard to apply any kind of golden cut to cleanly separate
signal and background. One useful property of the $s$-channel process
is that the final state particles decay from a pure $d$-wave (spin-2)
state.  This results in a distinctive angular separation $\Delta \phi$
of the final state muons~\cite{ang_corr}.\bigskip

What we are most interested in, though, is the theoretical basis of
the dimension-8 operator, \ie its derivation from the KK effective
theory. Its dimension $1/m^4$ coefficient arises partly from the
coupling and partly from the propagator structure
\begin{equation}
\s(s) = \frac{1}{\Mpl^2}\sum \frac{1}{s-\mkk^2} \; .
\end{equation}
It exhibits a sum over the KK tower with its typical small mass
spacing.  To replace it by an integral we need a quantity acting as
$\Delta \mkk$ which when sent to zero provides the Riemannian measure.
In a similar vein to the real mission case we can replace $\Mpl$ with
its definition in terms of $\Mstar$.  The factor of $r^n$ appearing in
the denominator is precisely $(\Delta \mkk)^2$.  This gives us an
integral over KK masses. Next, recalling that $\mkk \sim |\vec{n}|/r$
we realize that the only relevant coordinate in the KK state space is
the radial distance.  Therefore, it is possible to perform the angular
integration explicitly:
\begin{alignat}{5}
\s(s) &= \frac{1}{\Mstar^{2+n}}\int d^n \mkk \frac{1}{s-\mkk^2}.
\notag \\
      &=\frac{S_{n-1}}{\Mstar^{2+n}}\int d\mkk \frac{\mkk^{n-1}}{s-\mkk^2}
\label{eq:keyint}
\end{alignat}
The crucial point is that similar to eq.(\ref{eq:div_real}) this
integral is divergent for $n>1$, \ie for most phenomenologically
relevant scenarios. This divergence of the integrals in
eqs.(\ref{eq:div_real},\ref{eq:keyint}) is the unique feature of ADD
models which are based on an effective field theory of gravity. It can
only be cured by going a step beyond the effective theory and employ
some kind of UV completion of extra-dimensional gravity.\bigskip

Previously, we mentioned that summing over virtual graviton states is
similar to performing a loop-type integral.  However, unlike for
renormalizable gauge theories our effective theory of gravity has no
such thing as a well-defined counter term to absorb the UV divergence.
The remedy we use in this first discussion is to cut off the integral
explicitly at the limiting scale of our effective theory.  In the
spirit of an effective theory we study the leading terms in
$s/\cutoff^2$.
\begin{alignat}{5}
\s(s)
&= \frac{S_{n-1}}{\Mstar^{2+n}}\int_0^{\cutoff} d\mkk
   \frac{\mkk^{n-1}}{s-\mkk^2}
\notag \\
&= \frac{S_{n-1}}{\Mstar^4 (n-2)}
   \left(\frac{\cutoff}{\Mstar}\right)^{n-2}
   \left[1+\ope \left( \frac{s}{\cutoff^2} \right) \right]
\approx \frac{S_{n-1}}{n-2} \; \frac{1}{\Mstar^4}
\label{eq:conabsa}
\end{alignat}
where in the final line we identify $\Mstar \equiv \cutoff$, lacking
other reasonable options. Obviously, this relation should be
considered an order-of-magnitude estimate rather than an exact
relation valid to factors of two.  For simplicity this term is further
approximated in terms of a generic mass scale $\s=4\pi/M^4$ in the
literature~\cite{grw,gs}. A number of interesting properties of
virtual graviton exchange diagrams we can summarize.
\begin{itemize}
\item[--] The identification in eq.(\ref{eq:conabsa}) only
  parameterizes the effects of the transition scale between the well
  known linearized theory of gravitons and whatever new physics occurs
  above the scale $\Mstar$, assuming the effective theory captures the
  dominant effects.
\item[--] The generic $2\rightarrow 2$ $s$-channel amplitude $4\pi/M^4
  \cdot \mathcal{T}$ requires powers of the process energy in the numerator.
  Unitarity is violated at some $s$ signifying the breakdown of our
  effective theory.  For first collider predictions it suffices to
  make a hard cut on events with $\sqrt{s}>\cutoff$.  This as it will
  turn out poor approximation will be addressed later.
\item[--] The function $\s(s)$ including $\cutoff$ can be integrated
  analytically for a Wick rotated graviton propagator $1/(s+\mkk^2)$,
  but the interpretation of particles in an effective field theory
  will be lost beyond a leading approximation $s \gg \mkk^2$ or $s \ll
  \mkk^2$.
\end{itemize}
\bigskip

%----------------------------------------
\begin{figure}[t]
\begin{center}
\includegraphics[width=7cm]{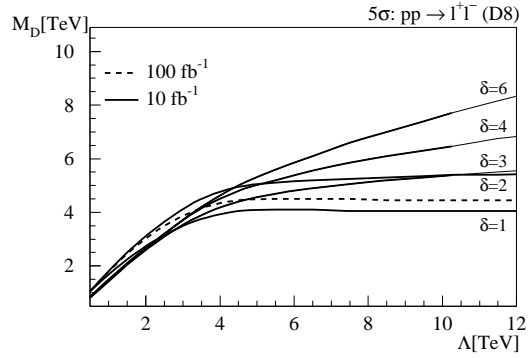}
\caption{$5\sigma$ discovery contours for discovery of extra
  dimensions via virtual graviton contributions to the Drell--Yan
  process. Figure from Ref.\cite{gps}.}
\label{fig:um23}	
\end{center}
\end{figure}
%----------------------------------------

Moving on to some results, the LHC reach in $M_D$ is given in
Fig.\ref{fig:um23}, again in the presence of a variable cutoff
$\cutoff$. As opposed to the real emission case shown in
Fig.~\ref{fig:ReEmm}, the result is clearly sensitive to $\cutoff$.
Only for $n=1$ this is not true, reflecting that $\int_0^{\infty} dm\;
\mkk^{-2}$ in fact converges.  Along a similar thread of thought, the
sensitivity becomes more pronounced with increasing $n$ as the
integral is more divergent.  The value of $\cutoff$ depends on the
details of the transition to UV behavior.  Phrased differently, the
LHC production cross section involving virtual graviton exchange is
seriously sensitive to the UV physics of quantum gravity.  This cutoff
dependence and the non-unitarity of gravitational scattering
amplitudes will be our main focus in the final section of these notes.

%%%%%%%%%%%%%%%%%%%%%%%%%%%%%%%%%%%%%%%%%%%%%%%%%%%%%%%%%%%%%%%%%%%%%%%
\section{Warped extra dimensions}
\label{sec:rs}

Our main focus thus far have been large flat dimensions as suggested
by the ADD model.  However, there is the alternative Randall--Sundrum
model~\cite{randall_sundrum,tasi_raman} with a wealth of
phenomenological applications~\cite{rs_constraints,rs_coll}.  It also
solves the hierarchy problem using an extra space dimension and
claiming that the fundamental Planck scale resides around the TeV
scale.  The mechanism which generates the large hierarchy between
$\Mstar$ and $\Mpl$ utilizes a spatially warped extra dimension.

For completeness, we also mention a third class of phenomenologically
relevant extra dimensional models, universal extra dimensions
UED~\cite{ued}. In this model all Standard Model particles exist in
the higher dimensional space, but the geometry is such that there is
an experimentally constrained mass gap between the ground state and
the first excited state.  At the LHC, we would for example expect to
see KK resonances of the gluon, \ie a massive color octet vector
particle. Since UED models do not provide a straightforward link to
quantum gravity effects we will not consider them in detail.\bigskip

For the Randall--Sundrum model, we compactify our 5th dimension $y$
on a $S^1/Z_2$ orbifold. $S^1$ simply means a circle, equivalent to
periodic boundary conditions. $S^1/Z_2$ means we map one half of this
circle on the other, so we really only look at half a circle with no
periodic boundary conditions, but two different branes at $y=0$ and
$y=b$.  The key observation now is that nobody can stop us from
postulating a 5-dimensional metric of the kind
\begin{equation}
 ds^2 = e^{- 2 k |y|} \eta_{\mu \nu} dx^\mu dx^\nu - d y^2 
 \qquad \Leftrightarrow \qquad 
 g_{AB}=\begin{pmatrix} e^{-2k|y|}\eta_{\mu\nu} & 0 \\
                        0 & \eta_{jk} \\
        \end{pmatrix}
\end{equation}
The metric in the four orthogonal directions to $y$ depends on $|y|$.
The absolute value appearing in $|y|$ corresponds to the $Z_2$
(orbifolding) as $S^1/Z_2$. When looking at our (3+1)-dimensional
brane we can take into account the warp factor $e^{-2k|y|}$ in two
ways (with some caveats):
\begin{enumerate}
\item Use $g_{\mu\nu}=\eta_{\mu\nu} e^{-2k|y|}$ everywhere, which is a
  pain but possible.
\item Replace $x^\mu$ in five dimensions by effective coordinates
  $e^{-k|y|} d\tilde{x}^\mu$ and $g_{\mu\nu}$ by
  $\tilde{g}_{\mu\nu}=\eta_{\mu\nu}$ where the tilde indicates
  4-dimensional variables.
\end{enumerate}
The second version means we shrink our effective 4-dimensional metric
along $y$ and forget about the curved space, because the warp factor
does not depend on $x^\mu$. The general--relativity action for
Newtonian gravity we can write in terms of the 5-dimensional
fundamental Planck scale $\Mrs$.  In our hand--waving argument we have
to transform the 5-dimensional Ricci scalar. Just looking at the mass
dimensions we see that $R$ has mass dimension two (or by looking at
the definition of $R$ we find space dimension minus two). This
suggests that the 4-dimensional Ricci scalar $\tilde{R}$ should
roughly scale like $x^{-2} \sim \tilde{x}^{-2} \exp{(+2k|y|)}$,
leading us to guess $R \sim \tilde{R} \exp{(+2k|y|)}$. The
Einstein--Hilbert action with separated $\tilde{x}$ and $y$ integrals
reads
\begin{alignat}{5}
 S &= -\frac{1}{2}\int_0^b \, dy \int \, d^4\tilde{x} \, 
	e^{-4k|y|} \; R\, \Mrs^3
        \notag\\
   &\sim -\frac{\Mrs^3}{2} \int_0^b \, dy \, e^{-2k|y|} \int 
 	\, d^4\tilde{x} \,
         \tilde{R}
        \notag\\
   &= -\frac{\Mrs^3}{4k} \left( 1 - e^{-2kb}
                         \right) \int \, d^4\tilde{x} \; \tilde{R}
       \notag\\
   &\sim - \frac{\Mrs^3}{4k} \int \, d^4\tilde{x} \tilde{R}
      \qquad &&\text{assuming $kb\gg 1$}
       \notag\\
   &\equiv -\frac{\Mpl^2}{2} \int \, d^4 \tilde{x} \; \tilde{R}
      &&\text{with} \qquad
    \Mpl^2 \sim \frac{\Mrs^3}{2k} \; .
\end{alignat}
In the last step we have applied the usual matching with 4-dimensional
Newtonian gravity. Note that this does yet not solve the hierarchy
problem because $\Mrs \sim k\sim \Mpl \sim 10^{19}$~GeV looks like the
most reasonable solution to the matching condition.\bigskip

Fortunately, this is not the whole story. Consider now the Standard
Model Lagrangian on the TeV brane ($y=b$) in the $\tilde{x}^\mu$
coordinates, \ie including the warp factor. To solve the hierarchy
problem, the scalar Higgs Lagrangian is obviously crucial
\begin{alignat}{3}
S_\text{SM} &= \int\, d^4\tilde{x}\, e^{-4kb}\, 
              \left[ (D_\mu H)^\dagger (D^\mu H)
                    - \lambda (H^\dagger H - v^2)^2
                    + ...
              \right]
\end{alignat}
From the Higgs mass term we see that we can rescale all Standard Model
fields and mass parameters --- in this case $H$ as well as $v$ --- by
the warp factor on the TeV brane $\exp{(-kb)}$. The same we have to do
for the space coordinate, as described above and for gauge fields
appearing in the covariant derivative. To get rid of the entire
pre-factor from the warped metric we need to absorb four powers of
$\exp{(-kb)}$ in each term contributing to the Standard Model
Lagrangian.

Four is a magic number in Lagrangians of renormalizable gauge
theories: it fixes the mass dimension of the Lagrangian. This means
that if we only consider contributions to $\lag_\text{SM}$ of mass
dimension four, we can simply rescale all Standard Model fields
according to their mass dimension:
\begin{alignat}{3}
 \tilde{H}     &= e^{-kb} H     &&\qquad\text{scalars}
         \notag\\
 \tilde{A}_\mu &= e^{-kb} A_\mu &&\qquad\text{or $\tilde{D}_\mu=e^{-kb}D_\mu$}
         \notag\\
 \tilde{\Psi}  &= e^{-3kb/2}\Psi&&\qquad\text{fermions}
\end{alignat}
which also means for all masses
\begin{alignat}{2}
 \tilde{m} & =e^{-kb}m \notag\\
 \tilde{v} &= e^{-kb}v 
\end{alignat}
Yukawa couplings as dimensionless parameters are not affected. If we
now assume $kb\sim 35$ we do solve the hierarchy problem:
\begin{equation}
  \tilde{v} \sim 0.1 \,e^{-kb} \; \Mpl \sim 0.1~\tev
\end{equation}
The fundamental Higgs mass and the fundamental Planck mass are indeed
of the same order, only the 4-dimensional Higgs mass (like all mass
scales on the TeV brane) appears smaller, because of the warped
geometry in the 5th dimension. In contrast, on the Planck brane with
its warp factor $\exp{(-k|y|) = 1}$ nothing has happened.\bigskip

Before we introduce gravitons as metric fluctuations into our RS
model, it turns out to be useful to rewrite the metric by rescaling
the 5th dimension $y \to z$ with
\begin{equation}
 ds^2 = e^{-A(z)} \left( g_{\mu\nu}dx^\mu dx^\nu - dz^2 \right)
\end{equation}
To simplify things we assume for the following brief discussion $y>0$.
This is obviously justified, as long as we limit our interest to the
TeV brane.  First, we define $A(z) = 2ky$ and rewrite the metric
\begin{equation}
 e^{-2ky} = e^{-A(z)} = \frac{1}{(1+k z)^2} 
   \qquad \Rightarrow \qquad
 dy = e^{-ky} \, dz = e^{-A(z)/2} \; dz
\end{equation}
The Planck brane at $y=0$ sits at $z=0$.  Assuming $k>0$ we find that
$y>0$ corresponds to $z>0$. The derivative indeed produces the 
 correct pre-factor of $dz^2$.\bigskip

To introduce tensor gravitons we expand the 4-dimensional part of the
metric:
\begin{equation}
  ds^2=e^{-A(z)} \left(\eta_{\mu\nu} \,
       + \, h_{\mu\nu}(x,z) \, dx^\mu dx^\nu \,
       - \, dz^2\right)
\end{equation}
Einstein's equation without sources but in the presence of $A(z)$ 
includes a linear term which does not look at all like an equation
of motion and which we therefore do not like. We can get rid of it
rescaling (as usual) $h_{\mu\nu}=e^{(2+n)kb/4}\widetilde{h}_{\mu\nu}$,
according to its bosonic mass dimension $[h]=m^{(2+n)/2}$.
This gives
\begin{alignat}{4}
  -\frac{1}{2}\p_C\p^Ch_{\mu\nu}+ \frac{2+n}{4} \, \p^CA \; \p_C h_{\mu\nu} =
  - \frac{1}{2}\p^C\p_C \widetilde{h}_{\mu\nu}
  + \left( \frac{9}{32}A'^2 - \frac{3}{8} A'' \right) \widetilde{h}_{\mu\nu}
&= 0 
\end{alignat}
as the equation of the motion for the rescaled graviton field
$\widetilde{h}_{\mu\nu}$. We can solve this equation of the motion by
separating variables
$\widetilde{h}_{\mu\nu}(x,z)=\widehat{h}_{\mu\nu}(x) \; \Phi(z)$ and
by giving mass to the tensor graviton solving $\p_\mu\p^\mu
\widehat{h}_{\mu\nu} = m^2 \widehat{h}_{\mu\nu}$. The equation of
motion
\begin{alignat}{5}
  - \p_z^2\Phi
  + \left( \frac{9}{16}A'^2-\frac{3}{4}A''
    \right) \Phi 
&= m^2 \Phi
\end{alignat}
is a Schr\"odinger-type equation for $\Phi$ with a potential term
\begin{equation}
 V(z) = \frac{9}{16} \frac{4k^2}{(k|z|+1)^2}
       +\frac{3}{4}\frac{2k^2}{(k|z|+1)^2}
      = \frac{15}{4}\frac{k^2}{(k|z|+1)^2}
\label{eq:rs_pot}
\end{equation}
This equation is first of all solved by the zero mode
\begin{alignat}{5}
     h^{(0)}_{\mu\nu} &=e^{+3A/4} \, \widetilde{h}_{\mu\nu}^{(0)}
                     = e^{+3A/4} \, \hat{h}_{\mu\nu}^{(0)}(x) \, \Phi^{(0)}(z)
                     \equiv \hat{h}_{\mu\nu}^{(0)}(x)
\end{alignat}
which in terms of the 5th coordinate $y$ means
$\Phi^{(0)}(y)=e^{-3k|y|/4}=e^{-3kb/4}$ on our TeV brane. So indeed,
gravity on the TeV brane is weak because of the exponentially
suppressed wave-function overlap.\bigskip 

Again, using the Schr\"odinger-type equation with $V(z)$ as given in
eq.(\ref{eq:rs_pot}) we can compute the KK graviton masses in our
4-dimensional effective theory
The boundary conditions $\p_z h_{\mu\nu} = 0$ on the branes are given
by the orbifold identification $y\rightarrow\,-y$ and assuming $z>0$.
On the two different branes we find
\begin{equation}
 \p_z^2\Phi=-\frac{3}{2}k\Phi \bigg|_\text{Planck}\qquad
 \p_z^2\Phi=-\frac{3}{2}\frac{k}{kz+1}\Phi \bigg|_\text{TeV}
\end{equation}
The solution of the equation of motion can now be expressed in terms
of Bessel functions, which are numbered by an index which corresponds
to the mass introduced above:
\begin{equation}
 \Phi_m(z) = \frac{1}{\sqrt{kz+1}}
             \left[ a_m \, Y_2\left(m\left(z+\frac{1}{k}\right)\right)
                   +b_m \, J_2\left(m\left(z+\frac{1}{k}\right)\right)
             \right]
\label{eq:rs_bessel}
\end{equation}
The masses of these modes are given in terms of the roots of the
Bessel function $J_1(x_j)=0$ for $j=1,2,3,4\cdots$
\begin{alignat}{2}
m_j=x_j\,k\,e^{-kb} \sim x_j\,\Mpl\,e^{-kb} \sim x_j \, \tev
\qquad \text{with} \quad x_j=3.8,7.0,10.2,16.5 \cdots
\end{alignat}
This means that the KK excitations in the Randall--Sundrum model with
one warped extra dimensions are almost, but not quite equally
spaced. If we remember that we can choose $kb\sim 35$ to solve the
hierarchy problem they are in the TeV range, \ie in contrast to the
ADD model not only resolvable by the LHC experiments put most likely
out of reach beyond $j=1$.\bigskip

In the last step we need to compute the coupling strength of these
heavy KK gravitons to matter, like quarks or gluons.  Remember that in
the ADD case we find tiny Planck-suppressed couplings for each
individual KK graviton, which corresponds to an inverse-TeV-scale
coupling once we integrate over the KK tower. For the warped model the
relative coupling strengths on the Planck brane and on the TeV brane
are approximately given by the ratio of the wave function
overlaps. While the zero-mode graviton has to be strongly localized on
the Planck brane, to explain the weakness of Newtonian graviton the
TeV brane, the KK gravitons do not have strongly peaked wave functions
in the additional dimension. From eq.(\ref{eq:rs_bessel}) we can read
off the ratio of wave functions --- assuming that the Bessel functions
with their normalized arguments will not make a big difference
\begin{equation}
  \frac{\Phi(z)\big|_\text{TeV}}{\Phi(z)\big|_\text{Planck}} 
  \sim \frac{\sqrt{kz+1}\big|_\text{Planck}}{\sqrt{kz+1}\big|_\text{TeV}} 
  \sim \frac{1}{e^{kb/2}}
  \sim \frac{1 \tev}{\Mpl}
\end{equation}
The coupling of the KK states is given by the left-hand side of
Einstein's equations which enters the Lagrangian just as for the large
extra dimensions. We have to distinguish between the flat zero mode
with un-suppressed wave function overlap and the KK modes with this
ratio of wave functions
\begin{equation}
\lag \sim 
  \frac{1}{M_\text{Planck}} T^{\mu\nu} h^{(0)}_{\mu\nu}
+ \frac{1}{M_\text{Planck} e^{-kb}} T^{\mu\nu} \sum h_{\mu\nu}^{(m)}
\end{equation}
We see that the heavy KK gravitons indeed couple with TeV-scale
gravitational strength and can be produced at colliders in sufficient
numbers, provided they are not too heavy.  Similarly to the flat extra
dimensions, the couplings of the different KK excitations are
(approximately) universal. Remembering the way the effective theory of
gravity breaks down in ADD models we see that integrating over an
ultraviolet regime of a KK tower of gravitons is not a problem in RS
models. However, in the large-energy limit we do find that first of
all scattering amplitudes computed in the effective RS model violate
unitarity, and that secondly in the ultraviolet regime the graviton
widths become large, which means the effective KK picture becomes
inconsistent. Both of these reasons again ultimately require an
ultraviolet completion of gravity.\bigskip

Obviously, this is phenomenologically very different from the flat
(ADD) extra dimensions. For warped extra dimensions we will not
produce a tightly spaced KK tower, but for example distinct heavy
$s$-channel excitations. One advantage of such a scenario is of course
that we can measure things like the KK masses and spins at colliders
directly~\cite{ang_corr}. The disadvantage for phenomenology is that
such resonance searches are boring and can be mapped on $Z'$
searches~\cite{tevatron_rs,zprime} one-to-one.\bigskip

%----------------------------------------------------------------------
\section{Ultraviolet completions}
\label{sec:uv}

As explicitly seen in the last two sections, the effective field
theory description of extra-dimensional gravity breaks down once the
LHC energies approach the range of the fundamental Planck scale. This
feature is expected --- if a coupling constant has a negative mass
dimension the relevant scale in the denominator has to be cancelled by
an energy in the numerator. Once this ratio of the typical energy over
the Planck scale becomes large gravity appears to become strongly
interacting and will eventually encounter ultraviolet poles. These
poles cannot be absorbed by our usual perturbative renormalization,
which means we cannot meaningfully quantize gravity without an
additional modification of this ultraviolet behavior.

We know several possible modifications of this dangerous ultraviolet
behavior~\cite{weinberg_new}: most well known, string theory includes
its own fundamental scale $M_S$ which is related to a finite size of
its underlying objects.  Such a minimum length acts as a an
ultraviolet cutoff in the energy, which regularizes all observables
described by the gravitational
interaction~\cite{veneziano,lhc_strings}. An alternative approach
which avoids any ad-hoc introduction of radically different physics
above some energy scale is based on the ultraviolet behavior of
gravity itself~\cite{weinberg}: the asymptotic safety scenario is
based on the observation that the gravitational coupling develops an
ultraviolet fixed point which avoids the ultraviolet divergences
naively derived from power
counting~\cite{fp_early,ergf,reuter_misc,percacci_misc,daniel_misc}.\bigskip

Going back to large flat extra dimensions and the ADD model the
divergence of the integral representing the sum over KK states (for
example shown in eq.(\ref{eq:keyint})) is a major phenomenological
problem.  The unphysical cutoff dependence seriously weakens our
ability to make precise LHC predictions or interpret possible LHC
results.  Furthermore, at a conceptual level this break-down of the KK
effective theory already at LHC energies insinuates the immediate need
for a more complete description of gravity~\cite{lp,tom_joanne}.
There are a number of effective-theory proposals which side-step this
complication by defining the integral in a cutoff independent scheme.
One such treatment relates to the eikonal approximation to the
$2\rightarrow 2$ process~\cite{eikonal}, another involves introducing
a finite brane thickness~\cite{thick_brane}.  In this example, the
gravitational coupling is exponentially suppressed above $\Mstar$ by a
brane rebound effect.  This is applied to the case of high energy
cosmic rays interacting via KK graviton exchange~\cite{pluemi}.
However, none of these models offer a compelling UV completion for the
KK effective theory.\bigskip

The fundamental deficiency in the description of extra-dimensional
gravity we discuss based on two approaches: in the context of string
theory, one initial motivation for the ADD model~\cite{early}, the
expectation is for string Regge resonances to appear above the string
scale. On the other hand, following our original motivation for the
ADD model --- its minimal structure with no additional states and a
very straightforward geometry --- we will focus on a UV completion
based on the observation of asymptotic
safety~\cite{weinberg,fp_extrad}.  This idea can be applied to LHC
phenomenology in ADD~\cite{lp} or RS~\cite{tom_joanne} models.

%----------------------------------------------------------------------
\subsection{String theory}

One possible ultraviolet completion of gravity could be string theory
with its finite minimum length scale regularizing the ultraviolet
behavior of transition amplitudes. For example, we can compute the
scattering $q \bar{q} \rightarrow \mu^+ \mu^-$ using open string
perturbation theory. Without tagging a certain vacuum with the
Standard Model as its low-energy limit, we can nevertheless construct
realistic string amplitudes for generic gauge and fermion fields.  The
first step is to restrict the (massless) fields to a D3 brane. Gauge
bosons are included by adding Chan-Paton factors $\lambda^a_{ij}$ at
the string endpoints.  For $i,j$ running from 1 to $N$ this implies
$N^2$ additional degrees of freedom, identical to the generators for
$U(N)$.  The Standard Model subgroups $SU(2)$ and $U(1)$ are thus
easily embedded.  The helicity amplitudes for $2\to 2$ scattering are
simply analytic functions in $s$, $t$ and $u$ together with the common
Veneziano amplitude~\cite{veneziano}
\begin{equation} 
 \s(s,t)=\frac{\Gamma(1-\alpha's)\Gamma(1-\alpha't)}
 			{\Gamma(1-\alpha's-\alpha't)}.
\end{equation}
in terms of the inverse string scale $\alpha' = 1/M_S^2$.  While we do
not exactly know the size of this scale it should lie between the weak
scale $v = 246$~GeV and the fundamental Planck scale $\Mstar$. For our
purposes it suffices to consider three distinct limits:
\begin{itemize}
\item[--] In the hard scattering limit $s\to\infty$ and for a fixed
  scattering angle (or equivalently fixed Mandelstam ratio of
  variables $t/s$) the amplitude behaves as
\begin{equation}
\s(s,t)\sim e^{-\alpha'(s\,\log s+t\,\log t)}
\end{equation}
  This can be seen by applying Stirling's approximation.  The physics
  is immediately apparent: due to the finite and dimensionful string
  scale $\alpha'=1/M_S^2$ all scattering amplitudes becomes weak in
  the UV. Unfortunately, this particular limit is not very useful for
  LHC phenomenology.

\item[--] The Regge limit for small angle high energy scattering in
  terms of Mandelstam variables means $s\to\infty$ with $t$ fixed.  In
  this limit the poles in the $\Gamma$ functions determine the
  structure: for $\sqrt{s}>M_S$ there appear single poles at negative
  integer arguments $1-s/M_S^2 = -(n+1)$ where $n=1,2,...$. These
  poles lie at $s=n M_S^2$, which tells us that string resonances
  appear as a tower of resonances in the $s$ channel.  Starting from
  the energies around $M_S$ this UV completion consists of a string of
  real particles with masses $\sqrt{n} M_S$.

\item[--] The leading corrections in $\alpha'$ valid for energies
  $\sqrt{s}$ below the string scale is
\begin{equation}
\frac{\Gamma(1-s/M_S^2) \; \Gamma(1-t/M_S^2)}{\Gamma(1-(s+t)/M_S^2)} 
= 1 - \frac{\pi^2}{6} \; \frac{st}{M_S^4} 
  + \ope\left( \frac{1}{M_S^6} \right).
\label{eq:Stringred} 
\end{equation}
  This form of the string corrections corresponds to our KK effective
  field theory, modulo a normalization factor which relates the two
  mass scales $M_S$ and $\Mpl$. Hence, this series in $M_S$ is not
  what we are interested as the UV completion of our theory.

\item[--] The physical behavior for scattering amplitudes above the
  string scale is a combination of Regge and hard scattering behavior.
  In other words, equally spaced string resonances together alongside
  exponential suppression, but at colliders the resonances should be
  the most visible effects.
\end{itemize}
\bigskip

So far, we have only considered the exchange of string resonances of
Standard Model gauge bosons, not graviton exchange.  The string theory
equivalent of our process generating the effective dimension-8
operator is the scattering of four open strings via the exchange of a
closed string.  This amplitude is insignificant compared to the string
excitations in the vicinity of the string scale, where
eq.(\ref{eq:Stringred}) is valid~\cite{lhc_strings}.  Most notably,
the KK mass integration is finite for all $n$ due to an exponential
suppression of similar origin as the hard scattering behavior noted
above.  For fields confined to a D3 brane this integral is
\begin{equation}
\s(s) \sim \int d^6 m \; 
      \frac{e^{\alpha'(s-m^2)/2}}{s-m^2} \; .
\end{equation}
An explicit exponential factor regularizing the $m$ integral also
appears in the modification using a finite brane
thickness~\cite{thick_brane}
\begin{equation}
\s(s) \sim \int d^6 m \; \frac{e^{-\alpha' m^2}}{s-m^2}  \; .
\end{equation}
These results are thus approximately equivalent in the low $\sqrt{s}$
region for the $m$ integration.  The finite brane thickness approach
still violates unitarity for large $s$. \bigskip

\begin{figure}[t]
\begin{center}
\includegraphics[width=7.3cm,]{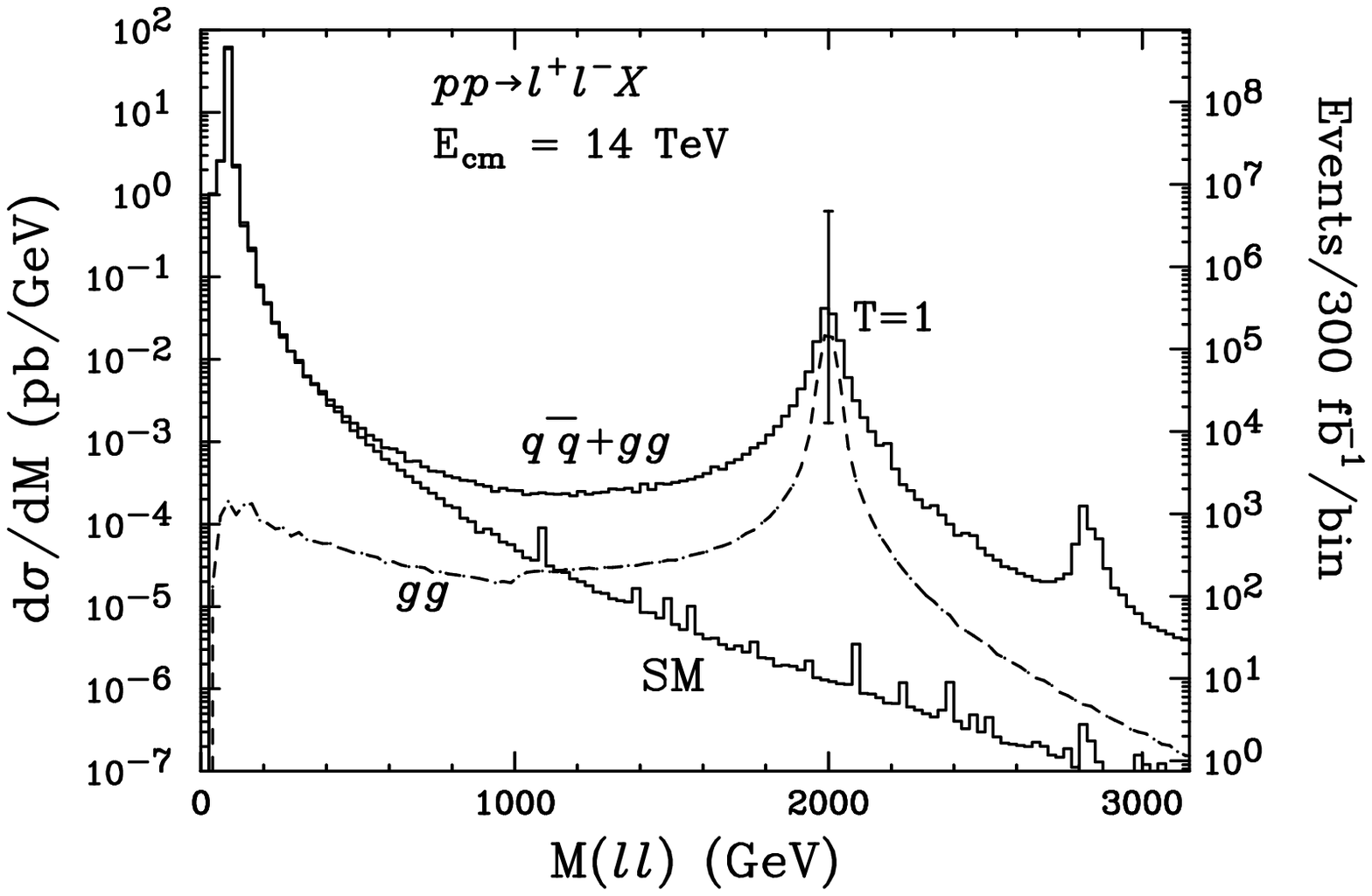}
\hspace*{5mm}
\raisebox{1mm}{\includegraphics[width=6.4cm,]{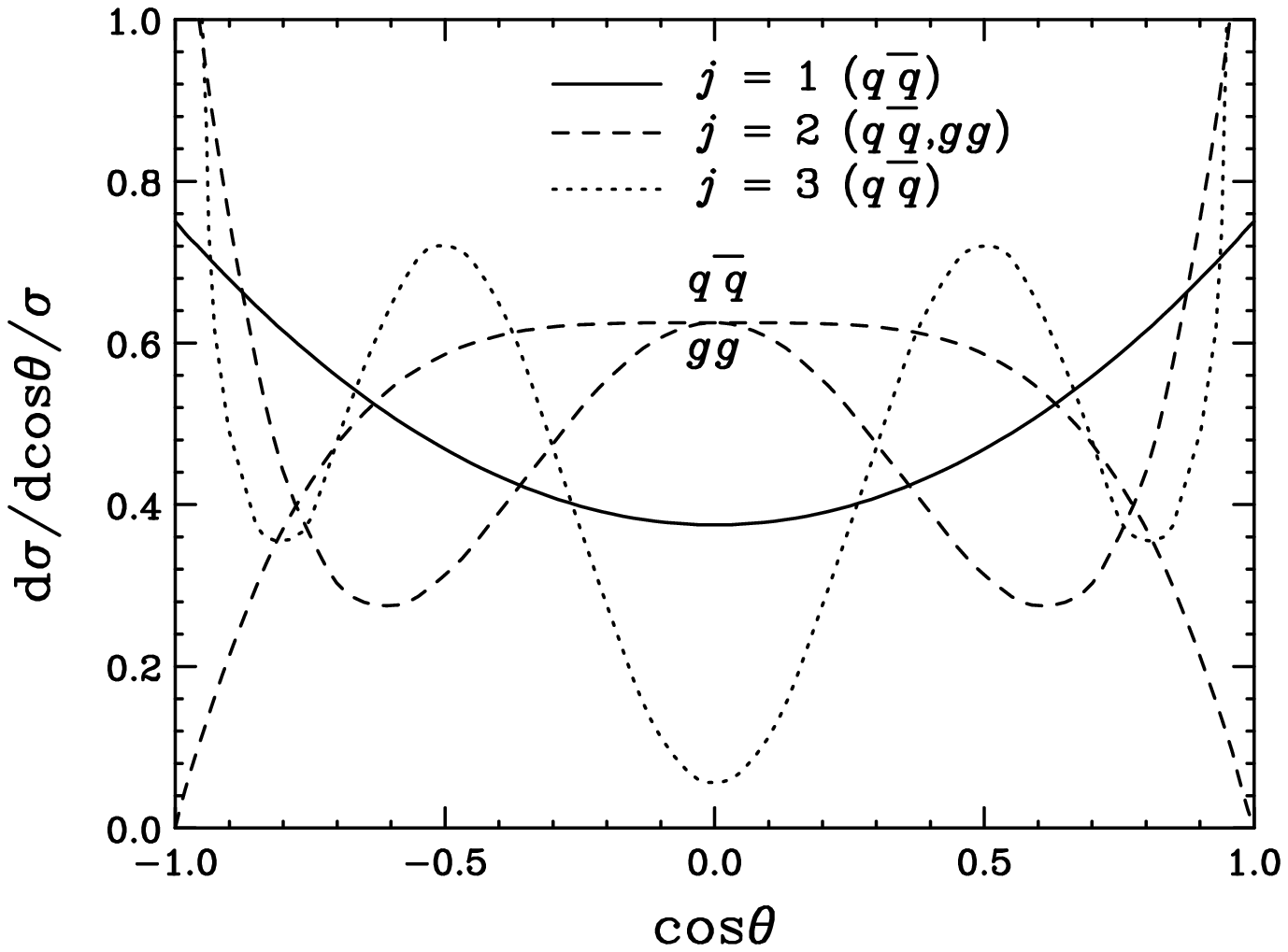}}
\caption{Left: invariant mass distribution for LHC dilepton
  production.  The parameter $T=1$ is the Chan-Paton number. Figure
  from Ref.\cite{lhc_strings}. Right: normalized angular
  distributions for $J=1,2,3$ resonances. Figure from
  Ref.~\cite{lhc_strings}.}
\label{fig:STRING}	
\end{center}
\end{figure}

Matching the string theory result with the effective theory amounts to
a matching between the string scale and the cutoff scale appearing in
the dimension-8 operator written as in eq.(\ref{eq:conabsa})
\begin{equation}
\frac{1}{M^4} \sim \frac{\pi^2}{32}\frac{g^4}{M_S^4}
\end{equation} 
This is the basis on which to argue that Regge string excitations are
the dominant process for high energy particle scattering.  Discovering
string degrees of freedom at the LHC is primarily concerned with
observing these resonances. An example for a possible distribution
plot is given in the left panel of Fig.\ref{fig:STRING}.  In addition,
the partial wave decomposition of a string Regge amplitude shows
superposition of different angular momentum state.  Each resonance is
degenerate with respect to the angular momentum number $j$.  This
degeneracy manifests itself in the angular correlations between final
states in $s$-channel processes (right panel of
Fig.\ref{fig:STRING}).\bigskip

%----------------------------------------
%----------------------------------------

As mentioned above, the issue with the naive phenomenology of RS
gravitons as well as string excitations is no experimental or
phenomenological challenge to the Tevatron or LHC communities. For
example the Tevatron experiments have been searching for (and ruling
out) heavy gauge bosons, for a long time.  So while a discovery of a
$Z'$ resonance at the LHC would trigger a great discussion of its
origin, including KK gravitons, string resonances, KK gauge bosons, or
simple heavy $Z'$ gauge bosons from an additional gauged $U(1)$
symmetry~\cite{zprime}, there is little to learn from such scenarios
at this stage. Most of the papers you will find on the topic have
not much to say about the generic structure and challenges of such
signatures at hadron colliders.

%----------------------------------------
\subsection{Fixed-point gravity}

If all hints concerning the asymptotic safety of gravity should hold
there is no need at all to alter the structure of gravity at high
energies --- gravity will simply be its own ultraviolet
completion~\cite{weinberg,fp_early,ergf} For our phenomenological
discussion we will only sketch some qualitative features of asymptotic
safety~\cite{fp_review,reuter_misc,percacci_misc,daniel_misc}.
Although most in this field is developed in four dimensions, the
results generalize in a straightforward manner~\cite{fp_extrad}.  This
allows us to splice together results from two seemingly disjoint
fields and consider asymptotic safety in extra dimensional models.
The major principles for asymptotic safety in gravity are
\begin{itemize}
\item[--] The metric carries the relevant degrees of freedom in both
  the classical and quantum regime.
\item[--] IR and UV physics lie on a single trajectory and are
  connected by the renormalization group flow.
\item[--] Relevant degrees of freedom are anti-screening.
\item[--] the UV behavior is determined by an interacting
  (non-gaussian) fixed point of the gravitational coupling.
\item[--] Residual interactions appear 2-dimensional
\end{itemize}
Evidence for asymptotic safety comes in many forms: the concept of
asymptotic safety or non-perturbative renormalizability was proposed
originally in 1980~\cite{weinberg}. The first hints that gravity might
have a UV fixed point were uncovered using a $2+\epsilon$ expansion
for the space-time dimensionality~\cite{two_plus_eps}.  Further
evidence was collected in the $1/N$ expansion where $N$ is the number
of matter fields coupled to gravity~\cite{fp_early}.  More modern
results use exact renormalization group methods~\cite{ergf}.  There
are a number of reviews of the subject~\cite{fp_review}, and on the
necessary non-perturbative techniques, namely the exact flow equation
for the effective average action or Wetterich
equation~\cite{wetterich_eq}.  Gravitational invariants including
$R^8$ and minimal coupling to matter are consistently included in flow
equations without destroying the fixed point~\cite{fp_review}.  More
recently, it has been shown that including invariants proportional to
divergences in perturbation theory do not give divergent results
non-perturbatively~\cite{include_divergences}.  Furthermore, there is
independent evidence for asymptotic safety coming from recent lattice
simulations, causal dynamical triangulations~\cite{cdt}.  Universal
quantities, like \eg critical exponents, computed using this method
agree non-trivially with results derived using renormalization group
methods.

The key point for our application is that in the UV the coupling
exhibits a finite fixed point behavior.  This ensures that the UV
behavior of the complete theory is dominated by fixed point scaling,
rendering all our computed transition rates weakly interacting at all
energy scales.\bigskip

It is useful to start again from the Einstein--Hilbert action to
calculate the scaling behavior for the gravitational coupling $G_N\sim
1/\Mpl^{2+n}$
\begin{equation}
\Gamma= \int d^d x \; \sqrt{g} \;
           \left\{ \frac{1}{16\pi G_N} (-R+2\Lambda_\text{cc}) 
                 + \ope(R^2)
                 + \lag_\text{matter}
                 + \lag_\text{gauge fixing}
                 + \lag_\text{ghost}
           \right\} \; .
\label{eq:effgravact}
\end{equation}
The gauge-fixing and ghost terms in the Lagrangian we will ignore in
the following.  The effective action eq.(\ref{eq:effgravact}) we
truncate to only the cosmological constant and Ricci scalar terms.  It
is necessary to include the cosmological constant because the flow in
$G_N$ is correlated with a flow in $\Lambda_\text{cc}$, so even if
$\Lambda_\text{cc}$ is set to zero at some point, quantum effects will
generate a non-zero value.

The dependence of dimensionless couplings on the energy scale is
relevant to understanding quantum effects and thus we define a
dimensionless Newton constant~\cite{ergf,percacci_misc,daniel_misc,
reuter_misc}
\begin{alignat}{5}
G_N \rightarrow \frac{G_N}{Z(\mu)}  \equiv G_N(\mu)
\equiv \frac{g(\mu)}{\mu^{2+n}}
\end{alignat}
not to be confused with the determinant of the metric $\sqrt{g}$,
which will not appear anymore in the following.  The Callan-Symanzik
equation for $g(\mu)$ can be derived in the standard way: first, we
note that as we vary the energy scale Newton's constant undergoes 
multiplicative renormalization.
In addition, we define $\eta=-\mu \, d\log Z(\mu)/d \mu$ as its
anomalous dimension. This anomalous dimension encodes how quantum
effects affect the scaling behavior of our theory.  Applying the
differential operator $\mu \, d/d\mu = d/\log \mu$ to the definition
of the dimensionless coupling we see that
\begin{equation}
\beta_g = \frac{d}{d \log \mu} \; g(\mu) = 
\left[ 2+n+\eta \right] \; g(\mu) \; .
\label{eq:canbeta}
\end{equation}
This is the exact beta function of Newton's constant.  Although it
looks innocuous enough at first glance, the parameter $\eta(g)$ will
in general contain contributions from all couplings in the Lagrangian,
not only the dimensionless Newton's constant.  However, one important
property is immediately apparent: for $g=0$ we have a perturbative
gaussian fixed point, \ie an IR fixed point which corresponds to
classical general relativity where we have not observed a running
gravitational coupling. Secondly, depending on the functional form of
$\eta(g)$ the prefactor $2+n+\eta(g)$ can vanish, giving rise to a a
non-gaussian fixed point $g_\star \ne 0$. The anomalous dimension at
this ultraviolet fixed point will take only integer values
\begin{equation}
\eta(g_\star)=-2-n.
\end{equation}
For two space-time dimensions the anomalous dimension vanishes in
which case the fixed point is at zero coupling and becomes a gaussian
fixed point, another manifestation of the perturbative
renormalizability of two-dimensional gravity.  It is tempting to think
that $G_N(\mu)$ will vanish at the UV fixed point so gravity is really
asymptotically free. However, Newton's dimensionful constant does not
have a physical meaning itself, and only when divided by an area does
it acquire significance. The corresponding dimensionless coupling does
not vanish, which means the correct statement is that in $4+n$
dimensions the theory with $g_\star \ne 0$ is still coupled, just not
ultraviolet divergent~\cite{percacci_review,fp_extrad}.  Using the
exact renormalization group flow equation we can compute the anomalous
dimensions~\cite{daniel_misc}, which depends on the shape of the UV
regulator. The beta function of the gravitational coupling becomes ---
neglecting the cosmological constant, which does not alter the
qualitative behavior of the system
\begin{alignat}{5}
\beta_g(g) &=\frac{(1-4 (4+n) g)(2+n)g}{1-(4+2n)g}\,
\qquad \qquad 
\eta(g) &=\frac{2(2+n)(6+n)\,g}{2(2+n)\,g-1}
\end{alignat}
Indeed, we observe the two fixed points: the IR fixed point
$\beta_g=0$ appears at zero coupling $g=0$ and the UV fixed point $g_*
= 1/4/(4+n)$ for an anomalous dimension of $\eta(g_*) = -2-n$.\bigskip

One way of interpreting the physical effects of the gravitational UV
fixed point is to modify the original calculations by defining a
running Newton's coupling and evaluate it at the energy scale given by
the respective process. This approach is in complete accordance with
the usual QCD calculations for high-energy colliders, based on a
running strong coupling. To derive a renormalization group equation
for the gravitational coupling we can integrate this form with respect
to a reference value $g_0 = g(\mu_0)$~\cite{lp}
\begin{alignat}{5}
 \log \frac{g(\mu)}{g_0} 
-\frac{6+n}{2(4+n)} 
 \log \frac{g(\mu) - g_*}{g_0 - g_*}
= \left( 2+n \right) \; \log \frac{\mu}{\mu_0} \; .
\label{eq:rge_explicit}
\end{alignat}
\bigskip

To motivate one method of including the ultraviolet fixed point in a
cross section calculation the renormalization group equation
eq.(\ref{eq:rge_explicit}) can be cast into the
form~\cite{bonanno_reuter,daniel_misc,tom_joanne}
\begin{alignat}{5}
  \frac{g(\mu)}{g_0} \left( \frac{g_0 - g_*}{g(\mu) - g_*} \right)^{\omega g_*}
= \left( \frac{\mu}{\mu_0} \right)^{2+n}
\qquad \text{with} \quad
\omega = \frac{6+n}{2 (4+n) g_*}
\end{alignat}
For $\omega \sim g_*$ which happens
to be a reasonable approximation this becomes simply
\begin{alignat}{5}
%  \frac{g(\mu)}{g_0} \frac{g_0 - g_*}{g(\mu) - g_*} 
%= \left( \frac{\mu}{\mu_0} \right)^{2+n}
%\notag \\
%  \frac{g(\mu)}{g(\mu) - g_*} 
%= \frac{g_0 \left( \dfrac{\mu}{\mu_0} \right)^{2+n}}{g_0 - g_*}
%\notag \\
%  g(\mu) ( g_0 - g_* ) 
%= g_0 \left( \dfrac{\mu}{\mu_0} \right)^{2+n} (g(\mu) - g_*)
%\notag \\
%  g(\mu) \left( g_0 - g_* - g_0 \left( \dfrac{\mu}{\mu_0} \right)^{2+n} \right) 
%= - g_0 \left( \dfrac{\mu}{\mu_0} \right)^{2+n} g_*
%\notag \\
%  g(\mu) \left( 1 - \frac{g_0}{g_*} + \frac{g_0}{g_*} \left( \dfrac{\mu}{\mu_0} \right)^{2+n} \right) 
%= g_0 \left( \dfrac{\mu}{\mu_0} \right)^{2+n} 
%\notag \\
  g(\mu) 
= \frac{g_0 \left( \dfrac{\mu}{\mu_0} \right)^{2+n}} 
  {1 - \dfrac{g_0}{g_*} + \dfrac{g_0}{g_*} \left( \dfrac{\mu}{\mu_0} \right)^{2+n}}
\end{alignat}
As a very rough check this formula reproduces the non-gaussian fixed
point values $g(\mu)=g_*$ for $\mu\rightarrow \infty$ as well as the
gaussian fixed point $g=0$ for $\mu\rightarrow 0$.  The dimensionful
coupling $G_N(\mu)$ which becomes Newton's constant $G_N(\mu_0=0)
\equiv G_N$ in the far infrared is accordingly given by
\begin{alignat}{5}
G_N(\mu) & =\frac{G_N(\mu_0)}{1+ \dfrac{G_N(\mu_0) \mu^{2+n}}{g_*} 
                               -\dfrac{G_N(\mu_0) \mu_0^{2+n}}{g_*}}
=\left[\frac{1}{G_N}+\frac{1}{g_*} \mu^{n+2}\right]^{-1} \; .
\end{alignat} 
\bigskip

The leading effects from the renormalization group running of the
gravitational coupling we can now include into a correction (form
factor) to the coupling which appears in the Lagrangian $G_N=
(\sqrt{8\pi} \Mstar)^{-2}$, \eg in eq.(\ref{eq:KKLAG})
\begin{equation}
\frac{1}{\Mstar^{2+n}} \; h^{MN}T_{MN}
\rightarrow 
\frac{1}{\Mstar^{2+n}} \;
\left[1+\frac{1}{8\pi g_*}\left(\frac{\mu^2}{\Mstar^2}\right)^{1+n/2}
\right]^{-1} \; h^{MN}T_{MN}
\equiv \frac{F(\mu^2)}{\Mstar^{2+n}} \; h^{MN}T_{MN}
\end{equation}
which carries through in the decomposition of $h_{MN}$ to the
4-dimensional field $G_{\mu\nu}^{n}$. At high energies $\mu \gg
\Mstar$ the form factor scales like $F(\mu^2) \propto
(\Mstar/\mu)^{2+n}$.  The factor $1/(8 \pi g_*)$ is an $\ope(1)$
parameter controlling the transition to fixed point scaling.  

As for any renormalization scale choice, there is an inherent
ambiguity where to choose the scale $\mu$. In QCD calculations this
scale dependence vanishes once we include arbitrarily high orders in
perturbation theory, which in our case will not help.  For a collider
process, the simplest choice is $\mu=\sqrt{s}$, in which case the form
factor only has a noticeable effect for center of mass energies close
to $\Mstar$, as one might expect.\bigskip

With this form factor the KK mass kernel in our virtual graviton
exchange amplitude can be written as
\begin{alignat}{5}
\s(s)&=\frac{1}{\Mstar^{2+n}}\int d^n \mkk \frac{1}{s-\mkk^2}
%\notag
%\\
&\quad \rightarrow \quad 
\left[1+\left(\frac{s}{\Mstar^2}\right)^{1+n/2}\right]^{-1}
\frac{1}{\Mstar^{2+n}}\int d^n \mkk \frac{1}{s-\mkk^2}.
\end{alignat}
Note that if we treat $\mkk$ and $\sqrt{s}$ as separate scales and
evaluate the form factor in terms of $\sqrt{s}$, the $\mkk$
integration still requires a cutoff. On the other hand, as far as the
$s$ integration is concerned, the form factor solves the unitarity
problem associated with graviton scattering amplitudes at the
LHC. This can be seen by power counting: the amplitude for graviton
production is proportional to $1/\Mpl^2$. Summing over the KK tower
replaces this factor with the fundamental Planck scale
$1/\Mstar^2$. In addition, the geometry factor from the integration
adds a factor $1/\Mstar^n$, which together gives the $1/\Mstar^{2+n}$
we observe for example in eq.(\ref{eq:KKTxsec}). The form factor
compensates this precisely with its UV scaling $F(s) \propto
(\Mstar/\sqrt{s})^{2+n}$. The only thing we have to ensure is that the
numerical factor $1/(8 \pi g_*)$ does not spoil this
counter-play~\cite{tom_joanne}.\bigskip

%----------------------------------------
\begin{figure}[t]
\begin{center}
\includegraphics[width=4.5cm,angle=90]{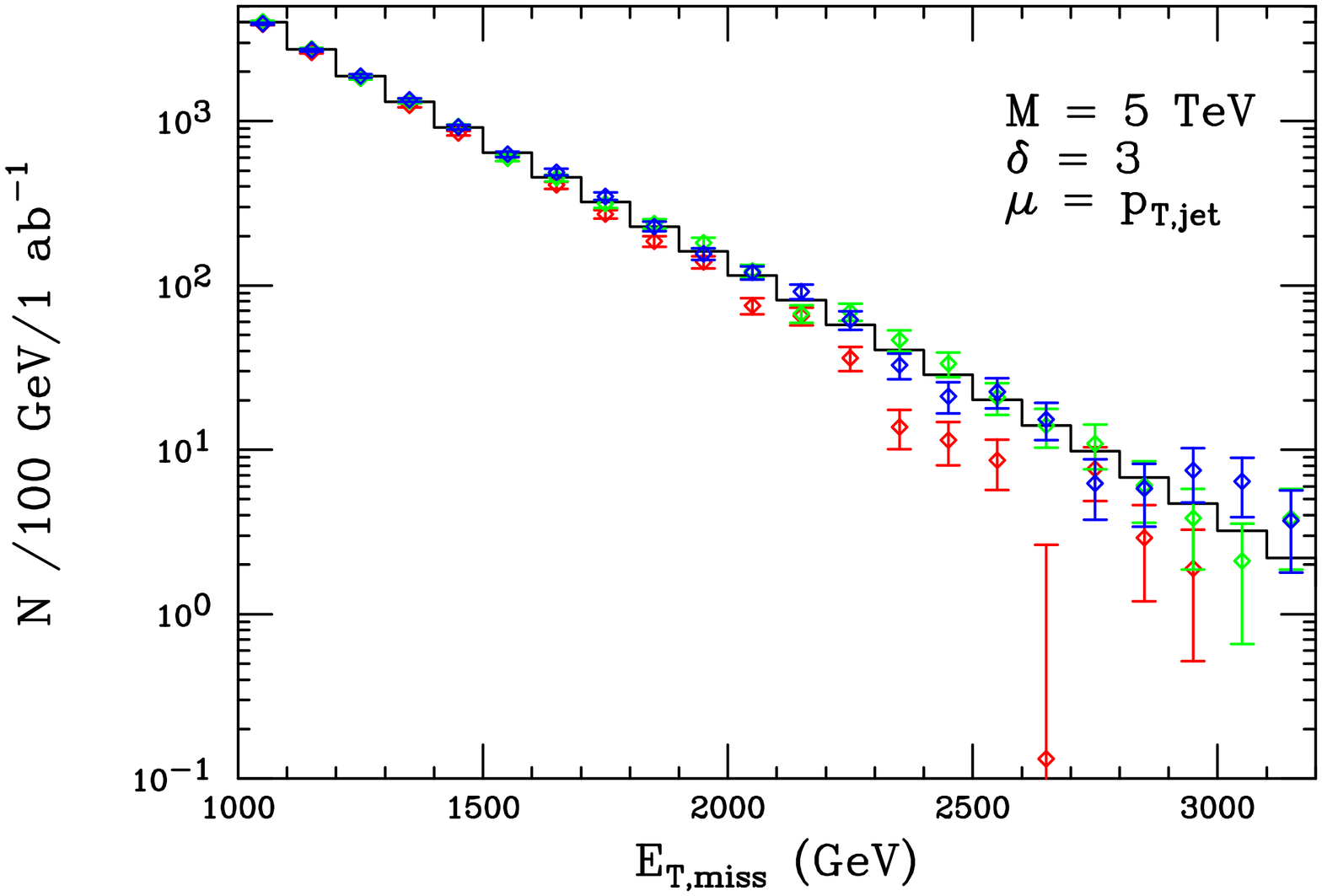}
\hspace*{10mm}
\includegraphics[width=4.5cm,angle=90]{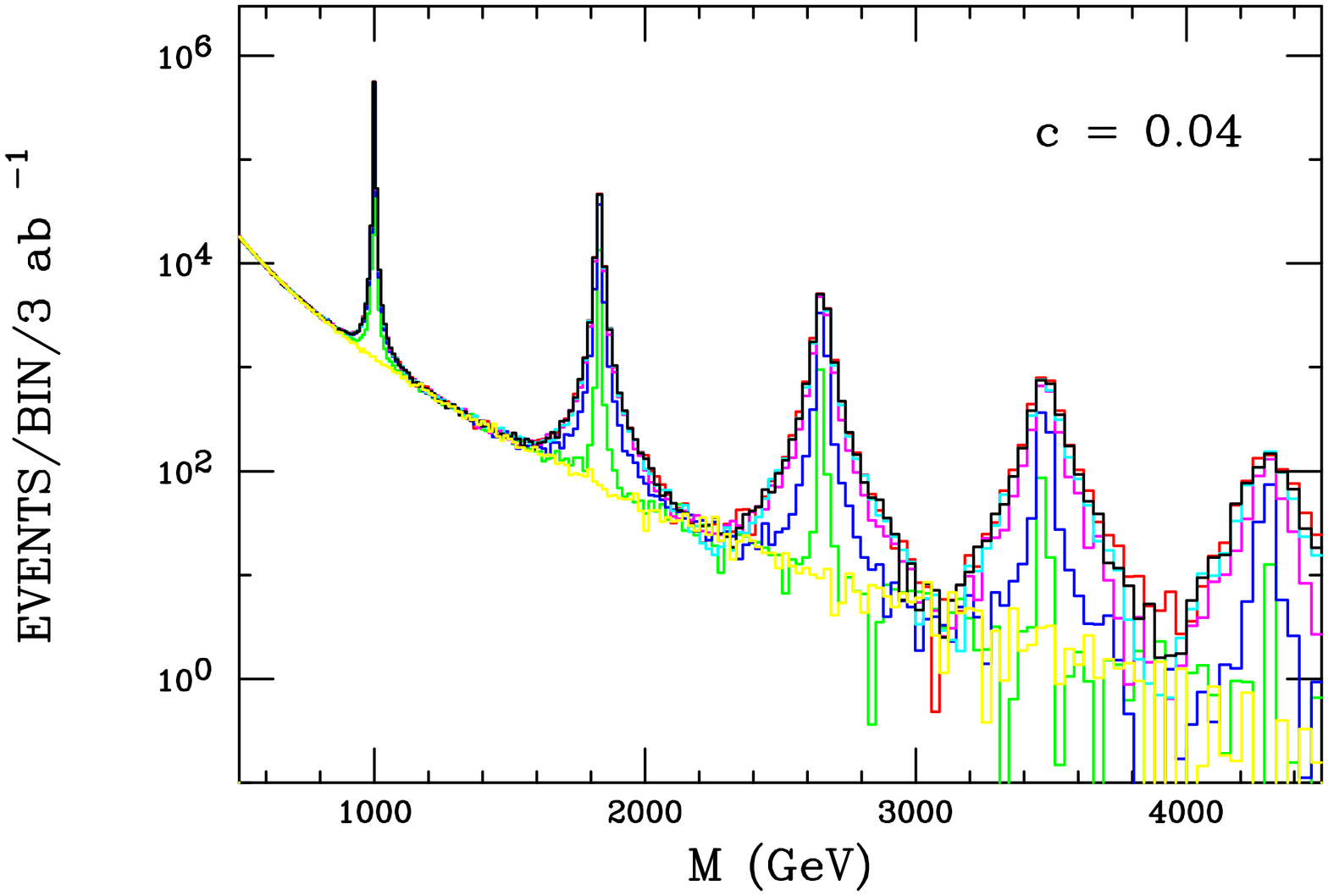}
\caption{Left: Missing transverse energy distribution
  for single jet + $\slashed{E}_T$ signal. The black 
  histogram represents standard ADD result for 
  1000 $\text{fb}^{-1}$ of LHC integrated luminosity. 
  Colored data points are for different parameterizations 
  of the fixed point cross-over.
  Right: Resonance RS graviton production at
  the LHC for a lightest KK graviton of $1\tev$
  Again, colored lines are for different parameterizations
  of the cross-over region.
  Both figures from Ref.\cite{tom_joanne}.}
\label{fig:HewRiz}	
\end{center}
\end{figure}
%----------------------------------------

Since $F(s)$ modifies the Planck scale or the gravitational coupling
in general, we can apply it to virtual as well as real graviton
emission, in the ADD model as well as in Randall--Sundrum models. One
example is the production of the first KK graviton excitations in
warped extra dimensions. As discussed before, those are single
particles produced for example in gluon fusion or quark-antiquark
scattering and decaying to jets or leptons.  The obvious effect of the
form factor is to reduce the number of gravitons produced at high
energies $\sqrt{\hat{s}}$.  For the Randall--Sundrum model we can see
this in the left panel of Fig.\ref{fig:HewRiz}.\bigskip

Virtual graviton processes in models with warped extra dimensions can
also be modified by fixed point effects.  The collider signal is
dominated by resonant graviton production, as opposed to the
unspecific KK tower in ADD models.  The form factor modifies the
coupling used when computing the width $\Gamma_j$ for the $j$-th RS
graviton.  For the production of one heavy state we run into the
convenient fact that there is only one scale in the process $\mu =
\sqrt{\hat{s}} = m_{j}$. The form factor becomes
\begin{equation}
F^{-1}=1+\left(\frac{m_{j}}{M e^{-\pi k r}}\right)^{3}.
\label{eq:RSWidth}
\end{equation}      
Without this form factor the width behaves like $\Gamma\sim
m_j^3/\Mrs^2$ and the resonance interpretation becomes less and less
valid at high energies (see Fig.\ref{fig:HewRiz}). Field theoretically
this is inconvenient since once the width becomes of the same order as
the mass spacing between modes the naive Breit-Wigner formalism breaks
down~\cite{bwwidth}. Including the form factor the signal if formed by
well defined resonances for higher masses, as shown in
Fig.\ref{fig:HewRiz}.\bigskip

%----------------------------------------
\begin{figure}[t]
\begin{center}
\includegraphics[width=7cm]{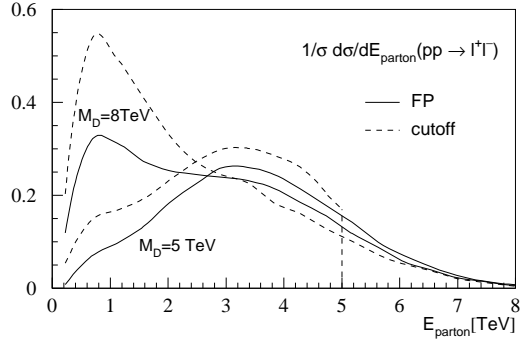}
\caption{Normalized distribution of the partonic energy (or $m_{\ell
    \ell}$) in the Drell--Yan channel for $n=3$.  The non-trivial
  shape difference between the $\Mstar=5\;\tev$ and $\Mstar=8\;\tev$
  is the result of interference effects. Figure from Ref.~\cite{lp}}
\label{fig:ERG2}	
\end{center}
\end{figure}
%----------------------------------------

An alternative (improved) method which is better suited for the case
of virtual gravitons is based directly on the form of the anomalous
dimension of the graviton eq.(\ref{eq:canbeta}).  It is motivated by
renormalization group techniques from condensed matter physics or
QCD~\cite{jan}.  In this picture, as usual there is a phase transition
occuring at a critical point with anomalous dimension $\eta_\star
=-2-n$ for some energy at or beyond the Planck scale.  At the critical
point, correlation functions are expected to scale by a function of
the anomalous dimension.  The momentum-space two point function
generically has the form $\Delta(p) \sim 1/(p^2)^{1-\eta/2}$, which
reproduces the classical result for small $\eta$.  In the vicinity of
the non-gaussian fixed point, this becomes
$1/(p^2)^{-(4+n)/2}$~\cite{daniel_optimized}.  The massive graviton
propagator in the fixed point region is then modified as
\begin{equation}
\frac{1}{s-\mkk^2}\rightarrow \frac{\mtrans^{2+n}}{(s-\mkk^2)^{(4+n)/2}}
\end{equation}
where the transition scale $\mtrans \sim \Mstar$ in the numerator
maintains the canonical dimensions for the propagator.  The graviton
kernel $\s(s)$ integrated over the entire $\mkk$ range becomes
\begin{equation}
\s(s) \rightarrow 
 \frac{1}{\Mstar^{2+n}} \int_0^{\mtrans} d^n \mkk \frac{1}{s-\mkk^2}
+\frac{1}{\Mstar^{2+n}} \int_{\mtrans}^\infty d^n \mkk 
 \frac{\mtrans^{2+n}}{(s-\mkk^2)^{(4+n)/2}}.
\end{equation}
This integral is finite to for all $n$ and the transition scale
$\mtrans$ parameterizes the crossover to fixed point scaling.  The
low-energy and high-energy contributions $\s_\text{IR/UV}(s)$ can be
easily calculated to leading order in $s/\mtrans^2$
\begin{alignat}{5}
\s_\text{UV} &= \frac{S_{n-1}}{4\Mstar^4}
\left(\frac{\mtrans}{\Mstar}\right)^{n-2}
\left[1+\ope\left( \frac{s}{\mtrans^2} \right) \right].
\notag \\
\s_\text{IR} &= \frac{S_{n-1}}{(n-2)\Mstar^4 }
\left(\frac{\mtrans}{\Mstar} \right)^{n-2}
\left[1+\ope\left( \frac{s}{\Lambda_T^2} \right) \right] \; .
\label{eq:irkernel}
\end{alignat}
The transition from the IR to UV scaling we for now treat as a
$\theta$-function.  The renormalization group predicts a smooth
transition, which can be modelled using a $\tanh{x}$ function.  The
following approximations though will not be sensitive to the
abruptness of the transitions, so we forgo implementing the smooth
transition for the remainder of this work.\bigskip

The combined IR and UV integral is given to leading order as
\begin{alignat}{5}
 \s=  \frac{S_{n-1}}{(n-2) \Mstar^4} \;
 \left(\frac{\mtrans}{\Mstar} \right)^{n-2} \left(1+\frac{n-2}
 {4}\right)
\end{alignat}
This result has several basic features:
\begin{itemize}
\item[--] We do not need any artificial cutoff scale.
\item[--] The result from the $\mkk$ integral has a small sensitivity
  to the precise value of the transition scale $\mtrans \sim\Mstar$,
  but including a more elaborate description of the transition region
  will remove this.
\item[--] For hadronic cross sections there is an additional integral
  over $s$ coming from the convolution with the parton distribution
  functions. Only in our leading-order approximation $\s$ is
  independent of $s$.
\item[--] For the full $\s_\text{UV}(s)$ perturbative unitarity is
  maintained by the large-$s$ behavior $\s_\text{UV}(s)\sim s^{-2}$
  given by dimensional analysis.  $\s_\text{UV}(s)$ and
  $\s_\text{IR}(s)$ do not naively match perfectly at the boundary
  $\sqrt{s}=\mtrans$, which requires a more careful treatment of this
  matching for the final numerical results.
\item[--] Phenomenologically, we do not expect resonance peaks or
  clearly distinctive features in the UV regime of graviton
  production. This feature is clearly different from the string theory
  completion.
\end{itemize}
\bigskip

%----------------------------------------
\begin{figure}[t]
\begin{center}
\includegraphics[width=15cm]{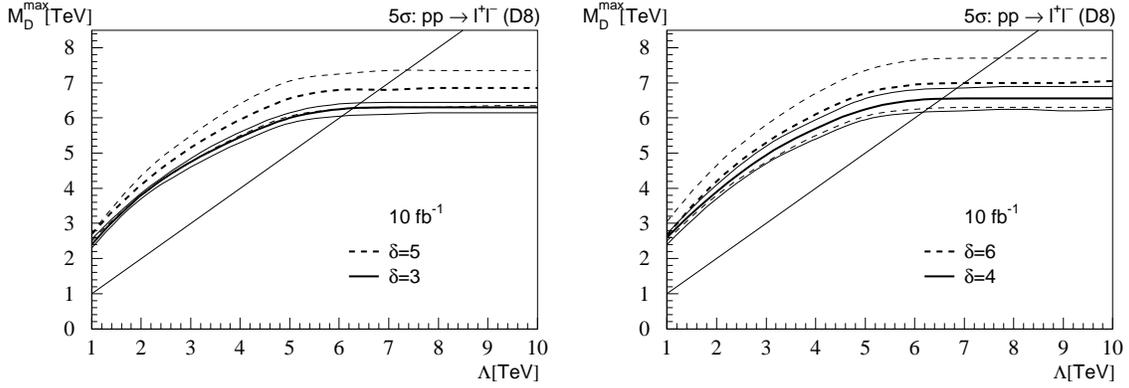}
\caption{5-$\sigma$ discovery contours at the LHC.  The solid diagonal
  line is for $\Mstar=\cutoff$.  The plateau in the discovery contours
  is the result of the UV fixed point (compare to
  Fig.\protect\ref{fig:um23}).}
\label{fig:ERG1}	
\end{center}
\end{figure}
%----------------------------------------

The anomalous dimension shift may also be implemented by evaluating
the Euclidean propagator
\begin{equation}
\s_\text{FP}(s)=
 \frac{1}{\Mstar^{2+n}} \int_0^{\mtrans} d^n \mkk \frac{1}{s+\mkk^2}
+\frac{1}{\Mstar^{n-2}} \int_{\mtrans}^\infty d^n \mkk 
      \frac{\mtrans^{2+n}}{(s+\mkk^2)^{(4+n)/2}}
\end{equation}
As we will see below, this agrees with the form factor provided that
we evaluate the coupling at a scale with $\mu=m$, reminiscent of the
RS form factor used to regulate the graviton width --- except that it
is expected to hold in general off-shell.  For large KK masses the
form factor behaves as
\begin{alignat}{5}
  \frac{1}{\bar{\Mstar}^{2+n}} d^n m \left[1+\frac{\omega}{8\pi}
  \left(\frac{\mkk^2}{\bar{\Mstar}^2}\right)^{1+n/2}\right]^{-1}
  \frac{1}{s+\mkk^2}
  \approx S_{n-1}\left[\frac{8\pi}{\omega}\right] \frac{dm}{\mkk^5} \; .
\end{alignat}
This is in agreement with the anomalous dimension shift which gives
\begin{alignat}{5}
\left( \frac{\mtrans}{\Mstar} \right)^{2+n} 
\frac{d^n m}{(s+\mkk^2)^{n/2+2}} 
\approx
S_{n-1}\left( \frac{\mtrans}{\Mstar} \right)^{2+n} \frac{dm}{\mkk^5} \; .
\end{alignat}
For the $s$ integration there is no clear agreement between the two
approaches.  The form factor in $s$ falls of much quicker than the
$1/s^2$ and gives a reduced cross section compared to our
estimate. This is expected since the second term in the expansion is
of the opposite sign. Note that by construction this Eucledian
argument avoids real particle poles in the graviton propagator and is
hence limited when comparing to an effective field theory.\bigskip

The search for extra dimensions at the LHC has two purposes.  The
first is to pin down the exact geometry of the non-visible space.  The
other and more interesting possibility is to ascertain a viable theory
of quantum gravity by probing energies beyond $\Mstar$.  The UV
completion to the KK integral can help us in both regards.  First of
all, the signal is enhanced by including the UV portion of the
integral.  In some cases this increase can be significant.  For
Drell--Yan production leading to final state muons the cross sections
are given in the following table.  Note that the LHC is expected to
provide $\sim 100~\text{fb}^{-1}$ per year at full luminosity, which
leads to a non-trivial event number for the following scenarios.

\bigskip

\begin{center}
\begin{tabular}{c|rrr|rrr} 
$\sigma [\text{fb}]$ & & $n=3$ &
                     & & $n=6$ &                \\[1mm]
\hline
$\Mstar$               & $2\;\tev$ & $5\;\tev$ & $8\;\tev$
                       & $2\;\tev$ & $5\;\tev$ & $8\;\tev$   \\
\hline
    $\s_{IR}$                   &  173 & 0.72 & 0.0204 &    66 & 0.28  & 0.008 \\
    $\s_{IR}+\s_{UV}$           &  408 & 1.24 & 0.0317 &   398 & 1.21  & 0.031 \\
\end{tabular}
\end{center}
\bigskip

In addition, the graviton kernel has a distinctive shape which
depends on the number of extra dimensions $n$.  In the $s$ channel at
lower partonic energies, the dominant interference term between
gravitons and $Z/\gamma$ imply a scaling with $\s \sim (n-1)$.  For
higher partonic energies the pure graviton amplitude is dominant and
the rate scales as $(n-1)^2$.  This fact is demonstrated in
Fig.\ref{fig:ERG2}.  The combined LHC reach in the virtual graviton
channel is given in Fig.\ref{fig:ERG1} and is mostly independent of
$n$ for the cases $n=3-6$.\bigskip

In this section we present a rough description of the effects of a
gravitational fixed point at the LHC. Many of the technical and
physical details are not worked out yet, because we are talking about
recent developments. However, we can convincingly argue that a
gravitational UV fixed point gives a consistent as well as complete
description of extra-dimensional observables at the LHC, clearly
distinguishable from alternative scenarios.

%----------------------------------------------------------------------
\section{Outlook}
\label{sec:outlook}

Large extra dimensions are a natural as well as truly minimal
extension of the Standard Model addressing the hierarchy problem.  If
they are realized in Nature, gravity effects become relevant at the
TeV scale and probing a viable theory of quantum gravity becomes an
experimental endeavor. Two generic approaches are either a free number
of large and flat extra dimensions (ADD model) or one extra dimension
with a warped metric (RS model). In both models only gravity with its
fundamental TeV-sized Planck scale propagates into the
extra-dimensional bulk, while our 4-dimensional Planck scale is a
derived observable. The relative size of the fundamental and
4-dimensional Planck scales can be derived by matching of the
effective 4-dimensional Kaluza--Klein effective theory. The main
phenomenological difference between the two models is the spacing of
the Kaluza--Klein masses, which is unobservably small in the ADD model
and of the order of the fundamental Planck scale in the RS
model.\bigskip

In particular for flat extra dimensions, the description by the KK
effective theory becomes increasingly unreliable once the experimental
energy reaches the TeV scale. The geometry factors from
compactification imply that any kind of observation will be dominated
by the respective UV tail of the graviton tower.  While real graviton
emission at the LHC is accidentally well described by an effective KK
theory (due to sharply falling parton luminosities) the prospect of
discovery via virtual graviton exchange depends on an unphysical
cutoff scheme regulating the sum over the states inside the KK
tower. For warped extra dimension a similar problem occurs in the
width of the KK gravitons, which is determined by the UV completion of
the model, \ie the quantum gravity regime.  Being a theoretical
ambiguity this generic UV cutoff dependence can act as a platform on
which to test candidate theories of quantum gravity.\bigskip

This means that the moment we probe extra dimensions at the LHC we
need to worry about the fundamental structure of gravity.  For
example, asymptotic safety or fixed-point gravity allows us to
consider gravity as a UV-complete theory, without introducing
additional states or ideas.  For the UV regime of gravity as probed at
the LHC it predicts a smooth fall-off in graviton amplitudes, clearly
different from resonances as expected by a string theory completion.
Once the LHC produces data we know there are a plethora of exciting
possibilities --- discerning among candidate theories of quantum
gravity, in the fortuitous scenario of large extra dimensions, ranks
among the most exciting prospects for the coming decade.\bigskip

%----------------------------------------------------------------------
\begin{center}
{\bf Acknowledgments}
\end{center} \smallskip

First and foremost, we are grateful to Daniel Litim for the exciting
collaboration on ultraviolet fixed points at the LHC and to the
BUSSTEPP school of 2006 for brokering this collaboration over great
real ales.  Daniel Litim also organized the school/workshop these
lecture notes are based on.  Moreover, TP would like to thank Gian
Giudice and Tao Han for their convincing arguments on the
attractiveness of extra dimensional theories.  EG would like to thank
the Institut f\"ur Theoretische Physik in Heidelberg for kind
hospitality during the writing of these notes, as well as SUPA and the
University of Heidelberg for their travel support.

%%%%%%%%%%%%%%%%%%%%%%%%%%%%%%%%%%%%%%%%%%%%%%%%%%%%%%%%%%%%%%%%%%%%%%%

\end{fmffile}
\end{document}